\documentclass[12pt]{article} 
\usepackage{amssymb,color,graphicx,here,float,appendix,amsmath} 

\newcommand{\nc}{\newcommand}

\nc{\la}{\lambda} \nc{\alf}{\alpha} \nc{\La}{\Lambda} \nc{\ze}{\zeta}
\nc{\tht}{\theta} \nc{\T}{\Theta} \nc{\be}{\beta}  \nc{\eps}{\epsilon} 
\nc{\ga}{\gamma}  \nc{\De}{\Delta}  \nc{\G}{\Gamma}  \nc{\vphi}{\varphi}
\nc{\de}{\delta} \nc{\si}{\sigma}  \nc{\ka}{\kappa}   \nc{\Si}{\Sigma} \nc{\rg}{{\rm g}}
\nc{\om}{\omega}  \nc{\qq}{\quad\quad}                \nc{\Om}{\Omega}
\nc{\nf}{\infty}   \nc{\dl}{\mathop{\smash{\cal L}}}  \nc{\black}{\rule{3mm}{3mm}}
\nc{\ra}{\rightarrow}    \nc{\ol}{\overline}        \nc{\und}{\underline} 
\nc{\beq}{\begin{equation}}  \nc{\eeq}{\end{equation}}  \nc{\pt}{\partial}  
   \nc{\dst}{\displaystyle}  \nc{\na}{\nabla} 
\nc{\nnb}{\nonumber}    \nc{\bs}{\backslash}        \nc{\mb}{\mathbb}   
\nc{\sn}{{\rm sn}\,} \nc{\cn}{{\rm cn}\,}     \nc{\dn}{{\rm dn}\,} \nc{\nin}{\noindent}
\nc{\ti}{\tilde}   \nc{\wti}{\widetilde}   \nc{\h}{\hat}  \nc{\wh}{\widehat}
\nc{\tpsi}{\wti{\psi}}   \nc{\tphi}{\wti{\phi}}  \nc{\tH}{\wti{H}} \nc{\Ai}{{\rm Ai}}

\newcounter{muni}
\newenvironment{remunerate}{\begin{list}{{\rm \arabic{muni}.}}
{\usecounter{muni}
\setlength{\leftmargin}{0pt}\setlength{\itemindent}{38pt}}}{\end{list}}

\nc{\brm}{\begin{remunerate}}   \nc{\erm}{\end{remunerate}}

 \newtheorem{lem}{Lemma} 
\newtheorem{nth}{Proposition}  \newtheorem{nTh}{Theorem}
\nc{\stg}{\mathop{\smash{*}}}
\nc{\st}{\mathop{\smash{\delta}}}
\nc{\barr}{\begin{array}}   \nc{\earr}{\end{array}}   \nc{\dg}{\dagger}
\nc{\mtvb}{\mathversion{bold}}   \nc{\mtvn}{\mathversion{normal}}  \nc{\F}{f_{\eps}}

\nc{\pf}{P_{\phi}} \nc{\pc}{P_{\chi}}

\begin{document} 

\begin{titlepage}

\date{\today}

\vspace{1cm}
\centerline{\bf\huge Global structure and geodesics}

\vspace{3mm}
\centerline{\huge\bf for Koenigs superintegrable systems }

\vskip 2.0truecm
\centerline{\large\bf  Galliano VALENT }

\vskip 2.0truecm
\centerline{ \it Laboratoire de Physique Math\'ematique de Provence}
\centerline{\it 19 bis Boulevard Emile Zola,  F-13100  Aix-en-Provence, France}
\nopagebreak

\vskip 2.5truecm

\begin{abstract} Starting from the framework defined by Matveev and Shevchishin we derive the local and the  global structure for the four types of super-integrable Koenigs metrics. These dynamical systems are always defined on non-compact manifolds, namely $\,{\mb R}^2\,$ and $\,{\mb H}^2$. The study of their geodesic  flows is made easier using their linear and quadratic integrals. Using Carter (or minimal) quantization we show that the formal superintegrability is preserved at the quantum level and in two cases, for which all of the geodesics are closed, it is even possible to compute the discrete spectrum of the quantum hamiltonian.
\end{abstract}

\end{titlepage}

\tableofcontents

\section*{Introduction}

In their quest for superintegrable systems defined on closed (compact without boundary) manifolds, Matveev and Shevchishin \cite{ms} have given a complete classification of all (local) Riemannian metrics on surfaces of revolution, namely 
\beq
\label{Ghhx}
G=\frac{dx^2+dy^2}{h_x^2},
\qq\qq
h=h(x),
\qq
h_x=\frac{dh}{dx},
\eeq
which have a superintegrable geodesic flow (whose Hamiltonian will henceforth be denoted by $H$), with  integrals $L=P_y$ and $Q$ respectively linear and cubic in momenta, {opening the way to the new field 
of {\em cubically} superintegrable models.} Let us first recall their main results.

They proved that if the metric $G$ is not of constant curvature, 
then ${\cal I}^3(G)$, the linear span 
of the cubic integrals, has dimension $4$ with a natural basis $P_y^3,P_y\,H,Q_1,Q_2$, and with the 
following structure. The map ${\cal L}:Q\to\{P_y,Q\}$ defines a linear endomorphism of 
${\cal I}^3(g)$ and one of the following possibilities hold:

\goodbreak

\brm
\item[(i)] 
${\cal L}$ has purely real eigenvalues $\pm \mu$ for some real $\mu>0$, then $\,Q_1$ and $\,Q_2$ 
are the corresponding eigenvectors.
\item[(ii)] 
${\cal L}$ has purely imaginary eigenvalues $\pm i\mu$ for some real $\mu>0$, then 
$Q_1 \pm iQ_2$ are the corresponding eigenvectors.
\item[(iii)] 
${\cal L}$ has the eigenvalue $\mu=0$ with one Jordan block of size $3$, in this case
\[\{L,Q_1\}=\frac{A_3}{2}\,L^3+A_1\,LH,\qq\qq \{L,Q_2\}=Q_1,\]
\erm
for some real constants $A_1$ and $A_3$. Superintegrability is then achieved provided the 
function $h$ be a solution of the following non-linear first-order differential equations:
\beq
\label{odes}
\barr{crcl}
(i)& \qq  h_x(A_0\,h_x^2+\mu^2\,A_0\,h^2-A_1\,h+A_2) &=& \dst  A_3\,\frac{\sin(\mu\,x)}{\mu}
+A_4\,\cos(\mu\,x)\\[4mm] (ii)&\qq  h_x(A_0\,h_x^2-\mu^2\,A_0\,h^2-A_1\,h+A_2) &=& \dst  A_3\,\frac{\sinh(\mu\,x)}{\mu}+A_4\,\cosh(\mu\,x)\\[4mm] 
(iii)& \qq h_x(A_0\,h_x^2-A_1\,h+A_2) &=& A_3\,x+A_4.
\earr
\eeq
We will denote case (i) as trigonometric, case (ii) as hyperbolic and case (iii) as affine.

The explicit form of the cubic integrals was given in all three cases. For instance,  
when $\mu \neq 0$, their structure is 
\beq
\label{F12}
Q_{1,2}=e^{\pm\mu y}\Big(
 a_0(x)\,P_x^3+a_1(x)\,P_x^2\,P_y+a_2(x)\,P_x\,P_y^2+a_3(x)\,P_y^3\Big),
\eeq
where the $a_i(x)$ are explicitly expressed in terms of $h$ and its derivatives, see \cite{ms}. The integration of these ODEs led to the explicit form of the metrics in local coordinates \cite{vds}, allowing to obtain all the globally defined systems on ${\mb S}^2$. Then, it was shown in \cite{Va3},  how to deduce easily their geodesics from the cubic integrals. 

However, as pointed out in \cite{ms}, the special case where $A_0=0$ is also of interest. In this special case the cubic integrals have the reducible structure $Q_{1,2}=P_y\,S_{1,2}$ and we are back to the SI systems first discovered by Koenigs \cite{Ko} where the extra integrals $(S_1,\,S_2)$ are now {\em quadratic} in the momenta, leading to a linear span ${\cal I}^2(g)$ of the quadratic integrals still of dimension 4 with basis
\[H\qq P_y^2 \qq S_1 \qq S_2.\]

The local structure of these systems has been thoroughly analyzed in the articles \cite{kk1} and \cite{kk2} with particular emphasis on the separability of the Hamilton-Jacobi and the Schrodinger equations. They also generalized Koenigs systems by computing some potentials $V(x,y)$ preserving superintegrability but we will restrict ourselves to the case of a potential $V(x)$ in order to preserve the Killing vector $\pt_y$. 

More recently, further potentials were derived in \cite{Ra}, while in \cite{kclm} with emphasis on the geodesics.  

The aims of this article are the following:
\brm 
\item To construct, starting from Matveev and Shevchishin setting for $A_0=0$, the local structure of the Koenigs models and to compare with Koenigs results. 
\item To determine, according to the values of the parameters defining each model, which ones are globally defined and on what manifold. We will exclude from our analysis the degenerate cases where the metrics have constant curvature.
\item For the globally defined  metrics, we will show how the superintegrability of their geodesic flow gives a direct access to  their geodesics.
\erm 

For the trigonometric case this is done in sections 2 to 4. For the hyperbolic case this is done in sections 
5 to 10. For the affine case this is done in sections 11 to 13. Section 14 is devoted to some concluding remarks.

\part{\bf\Large THE TRIGONOMETRIC CASE}

\setcounter{section}{0}
\section{Local structure}
Let us begin with the derivation of the metric and the quadratic integrals starting from Matveev and Schevchishin equations:

\begin{nth} The SI Konigs systems
\beq
{\cal I}_1=\{H,\ P_y,\ S_+\}\qq\qq {\cal I}_2=\{H,\ P_y,\ S_-\}\eeq
are given locally by
\beq\label{hamiL}
H=\frac{\sin^2 x}{2(1-\rho\,\cos x)}(P_x^2+P_y^2)\eeq
with
\beq\label{inti}
S_+=e^y\left(\sin x\,P_x\,P_y+\cos x\,P_y^2-\rho\,H\right), \quad     
S_-=e^{-y}\left(-\sin x\,P_x\,P_y+\cos x\,P_y^2-\rho\,H\right).\eeq
\end{nth}

\nin{\bf Proof:} In the ODE (\ref{odes})(i) we must take $A_0=0$. By a scaling of $x$ we can set 
$\mu=1$ and by a translation of $x$ we can take $A_4=0$. This ODE is easily integrated
\[-A_1 h h_x+A_2 h_x=A_3\,\sin x\qq\Longrightarrow\qq -\frac{A_1}{2}h^2+A_2 h=-A_3\,\cos x+A_4.\]
The scalar curvature being $\ R=2(h_x\,h_{xxx}-h_{xx}^2)\,$, it is constant for $A_1=0$. Hence $A_1$ cannot vanish so we can take $A_1=-2$ and $A_2=-2h_0$, leading to
\[h-h_0=\pm\sqrt{a_0+a_1\,\cos x}\qq\Longrightarrow\qq h_x=\mp\frac{a_1\,\sin x}{2\sqrt{a_0+a_1\,\cos x}},\]
with two constants $(a_0,\,a_1)\in{\mb R}^2$. Up to a global scaling we obtain for final metric
\[g=(1-\rho\,\cos x)\ \frac{dx^2+dy^2}{\sin^2 x} \qq\Longrightarrow\qq  H=\frac{\sin^2 x}{2(1-\rho\,\cos x)}(P_x^2+P_y^2).\] 
Transforming the formulas given in \cite{ms} we obtain the integrals (\ref{inti}).$\quad\Box$

Let us compare with Koenigs results\footnote{We stick to  Koenigs numbering which was modified in \cite{kk1},  \cite{kk2} and followers.}, given in \cite{Ko}{ p. 378}. His type I has for metric
\[g=\frac{a(e^w+e^{-w})+b}{(e^w-e^{-w})^2}\,du\,dv\qq\qq w=\frac{u-v}{2}.\]
Upon the change of coordinates $\,(u=ix+y,\,v=-ix+y)$, this metric becomes 
\[-\frac g4=\frac{2a\,\cos x+b}{\sin^2 x}(dx^2+dy^2)\] 
which, up to an overall scaling, is indeed (\ref{hamiL}). 

As shown in \cite{kk2}, keeping the same quadratic integrals (\ref{inti}), one may add the potential 
\beq\label{poti}
V(x)=\frac{\xi}{2(1-\rho\,\cos x)}\eeq 
still preserving the Killing vector $\pt_y$. 

Let us study the global structure of the type I Koenigs hamiltonian equipped with this potential. 

\section{Global structure}
It follows from: 

\begin{nth} The SI Koenigs systems
\beq
{\cal I}_1=\{H,\ P_y,\ S_+\}\qq\qq {\cal I}_2=\{H,\ P_y,\ S_-\}\eeq
with
\beq\label{hami}
H=\frac 1{2(1-\rho\,\cos x)}\Big(\sin^2 x\,(P_x^2+P_y^2)+\xi\Big)\eeq
such that
\[(x,y)\in\,(0,\,\pi)\times{\mb R} \qq\qq \rho\in(0,1)\qq\qq \xi\in\,{\mb R},\] as well as the quadratic integrals
\beq\label{Si}
S_+=e^y\left(\sin x\,P_x\,P_y+\cos x\,P_y^2-\rho\,H\right), \      
S_-=e^{-y}\left(-\sin x\,P_x\,P_y+\cos x\,P_y^2-\rho\,H\right).\eeq
are globally defined on the manifold $M \cong {\mb H}^2$. 
\end{nth}

\nin{\bf Proof:} In the metric induced by the hamiltonian (\ref{hami}) we will take $\,x\in(0,\pi)$ and $y\in{\mb R}$. We have to exclude the values $\rho=0,\,\pm 1$ for which the metric becomes of constant curvature. To be riemannian this metric requires $\rho\in (-1,0) \cup (0,1)$ but the change 
$x \to \pi-x$ allows to restrict $\rho$ to $(0,1)$.
 
To determine the nature of the manifold $M$ let us define the new coordinate
\[\chi=\ln\left(\tan\frac x2\right)\qq\qq x\in(0,\pi)\ \to\ \chi\in{\mb R}.\]
The metric becomes
\[g=(1+\rho\,\tanh\chi)\,(d\chi^2+\cosh^2\chi\,dy^2).\]
Recalling that the manifold ${\mb H}^2$ is embedded into ${\mb R}^3$ according to
\[x_1^2+x_2^2-x_3^2=-1\qq\qq (x_1,x_2)\in{\mb R}^2\qq x_3\geq 1\]
it was  shown in \cite{Va2} that if we take
\[x_1=\cosh\chi\,\sinh y\qq x_2=\sinh\chi \qq x_3=\cosh\chi\,\cosh y \qq\qq (\chi,y)\in{\mb R}^2\]
we have
\[d\chi^2+\cosh^2\chi\,dy^2=dx_1^2+dx_2^2-dx_3^2=g_0({\mb H}^2).\]
So we have obtained the relation
\[g=(1+\rho\,\tanh\chi)\ g_0({\mb H}^2).\]
Since the conformal factor is $C^{\nf}([0,+\nf))$ it follows that this metric is globally defined 
on a manifold $M$ diffeomorphic to ${\mb H}^2$.

To establish that the hamiltonian and the quadratic integrals are globally defined we have to use the generators (see \cite{Va2}):
\[\left\{\barr{l} M_1=\cosh y\,P_{\chi}-\tanh\chi\,\sinh y\,P_y\\[4mm]
M_2=P_y\\[4mm] 
M_3=-\sinh y\,P_{\chi}+\tanh\chi\,\cosh y\,P_y\earr\right.\]
with the $\,sl(2,{\mb R})$ Lie algebra
\[\{M_1,M_2\}=M_3\qq\qq \{M_2,M_3\}=-M_1 \qq\qq \{M_3,M_1\}=-M_2.\]
We obtain
\[H=\frac{\sqrt{1+x_2^2}}{2(\sqrt{1+x_2^2}+\rho\,x_2)}(M_1^2+M_2^2-M_3^2+\xi)\]
and
\[\frac{S_++S_-}{2}=M_2\,M_3+\rho\,\frac{x_1}{\sqrt{1+x_2^2}}\,H \qq\quad  
\frac{S_+-S_-}{2}=-M_1\,M_2+\rho\,\frac{x_3}{\sqrt{1+x_2^2}}\,H \]
which concludes the proof.$\quad\Box$

For future use let us point out the following useful relation
\beq\label{alphai}
\frac{(S_++S_-)}{2}\,\cosh y-\frac{(S_+-S_-)}{2}\,\sinh y=L^2\,\cos x-\rho\,E.
\eeq

Since the metric considered here is complete, by Hopf-Rinow theorem it is also geodesically complete. Let us now determine the explicit form of the geodesics.

\section{Geodesics}
As shown in \cite{Va3} the determination of the geodesics equations is quite easy for SI systems: they just follow from the non-linear integrals. The following points should be taken into account for all the subsequent discussions:

\brm
\item We will consider the invariant tori 
\[H=E\in\,{\mb R},\qq\qq P_y=L>0,\]
so that the hamiltonian (\ref{hami}) gives 
\beq
P_x^2=\frac{2E(1-\rho\,\cos x)-\xi}{\sin^2 x}-L^2.\eeq
\item To determine the geodesic equation  $y(x)$ we will always take an initial condition for which $y=0$. The most general case is merely obtained by the substitution $y\to\,y-y_0$, where $y_0\in\,{\mb R}$, due to the invariance of the metric under a translation of $y$.
\item The discrete symmetry $y\,\to\,-y$ shows that if $y(x)$ is a geodesic then $-y(x)$ must be also a geodesic.
\erm

We will begin with the geodesics of vanishing energy.

\begin{nth} For $E=0$ we have the following equations for the geodesics:
\beq\label{geodi1}\barr{ccll}
(a)\quad & -1<\xi<0 & \qq \left\{\barr{ll}\dst 
\cosh y=\frac{\cos x}{\cos x_*}  & \qq x\in(0,x_*)
\\[5mm] \dst \cosh y=\frac{\cos(\pi-x)}{\cos x_*}  & \qq  x\in(\pi-x_*,\pi)\earr\right.\\[1cm]
(b)\quad & \xi<-1 & \qq\dst  \eps\,\sinh y_{\eps}=\frac{\cos x} {\sinh\tht}  & \hspace{-3.4cm} x\in(0,\pi)\qq \eps=\pm 1\\[5mm] 
(c)\quad & \xi=-1 & \hspace{1.8cm} \dst e^{\eps\,y_{\eps}}=|\cos x| & \hspace{-3.2cm} x\in(0,\pi/2)\cup(\pi/2,\pi)\earr
\eeq
where 
\beq\label{xstari1}
\sinh\tht=\sqrt{|\xi|-1}\qq\qq \cos x_*=\sqrt{1-|\xi|}.\eeq
\end{nth}

\nin{\bf Proof:} Since we have
\[P_x^2=-\frac{\xi}{\sin^2 x}-L^2\]
we can set $L=1$ and we are left with a single parameter $\xi$. The positivity of $P_x^2$ requires $\xi<0$. 

The function $P_x^2$ has, for $x=\pi/2$, a minimum $p_*^2=|\xi|-1$. 
So if $\xi<-1$ then $p_*^2>0$ and the geodesic is defined for $x\in(0,\pi)$. We have
\[\sin x\,P_x=\eps\,\sqrt{\sinh^2\tht+\cos^2 x}\qq \eps=\pm 1,\]
where $\eps$ is the sign of the velocity. Taking for initial conditions $(x=\pi/2,\,y=0)$  we obtain $S_{\pm}=\pm\eps\,\sinh\tht$ and 
using relation (\ref{alphai}) we deduce (\ref{geodi1})(b).

If we have $-1<\xi<0$ then $p_*^2<0$ and the geodesics are defined either for $x\in(0,x_*)$ or for $x\in(\pi-x_*,\pi)$. In the first case the choice of the initial 
conditions $(x=x_*,\,y=0)$ gives $S_+=S_-=\cos x_*$. Using relation (\ref{alphai}) 
we obtain the first part of (\ref{geodi1})(a). The second part is merely obtained by the substitution $\,x\,\to\,\pi-x$. 

If we have $\xi=-1$ it is safer to use Hamilton equations
\[\dot{x}=\frac{\sin x\,P_x}{1-\rho\,\cos x}\qq 
\dot{y}=\frac{\sin x}{1-\rho\,\cos x}\qq\Longrightarrow\qq y'=\frac 1{P_x}=\eps\,\frac{\sin x}{|\cos x|}\]
which gives (\ref{geodi1})(c) if one takes as initial conditions $(x=0,\,y=0)$.
$\quad\Box$

\vspace{2mm}
\nin{\bf Remarks:}
\brm
\item Notice the very special case $\xi=-1$ for which 
the geodesics are asymptotes to $x=\pi/2$, where the velocity $\dot{x}$ vanishes. 
\item As we have seen, the quadratic conservation laws give quite easily the geodesic equations, except in case (c) for which they are degenerate.
\erm 

Having settled the zero energy case, let us define two new parameters $\si$ and $\eta$ by
\[\si=\frac 1{\rho}\left(\frac{\xi}{2\,E}-1\right) \qq\qq \eta=\frac{\rho\,E}{L^2}\]
which allow to write
\beq \label{Px2i}
\frac{P_x^2}{L^2}=-2\eta\,\frac{(\si+\cos x)}{\sin^2 x}-1 \qq\qq 
\frac{(P_x^2)'}{L^2}=2\eta\frac{(\cos^2 x+2\si \cos x+1)}{\sin^3 x}.\eeq

Let us consider first the geodesics with positive energy:

\begin{nth} For $\eta>0$ and $-1<\si<1$ the geodesic has for equation 
\beq\label{Epos}
\cosh y=\frac{\eta-\cos x}{\sqrt{\eta^2+2\si\eta+1}}  \qq \qq x\in\,(x_*,\pi)
\eeq
where $x_*$ is determined from
\beq\label{xsti}
\cos x_*=\eta-\sqrt{\eta^2+2\si\eta+1}.\eeq
\end{nth}

\nin{\bf Proof:}  If $\si\geq 1$ then $P_x^2$ is negative and there is no geodesic. If $-1<\si<1$ then $P_x^2$ increases monotonically from $-\nf$ to $+\nf$ so it vanishes for $x=x_*$ given by (\ref{xsti}), 
and the geodesic is defined for $x\in\,(x_*,\pi).$ 
The relation (\ref{alphai}), taking for initial conditions $(x=x_*,\,y=0)$, we have
\[(\cos x_*-\eta)\,\cosh y=\cos x-\eta\]
which gives (\ref{Epos}). $\quad\Box$

On the following figure some geodesics are drawn:

\begin{figure}[H]
\centering
\includegraphics[scale=0.6]{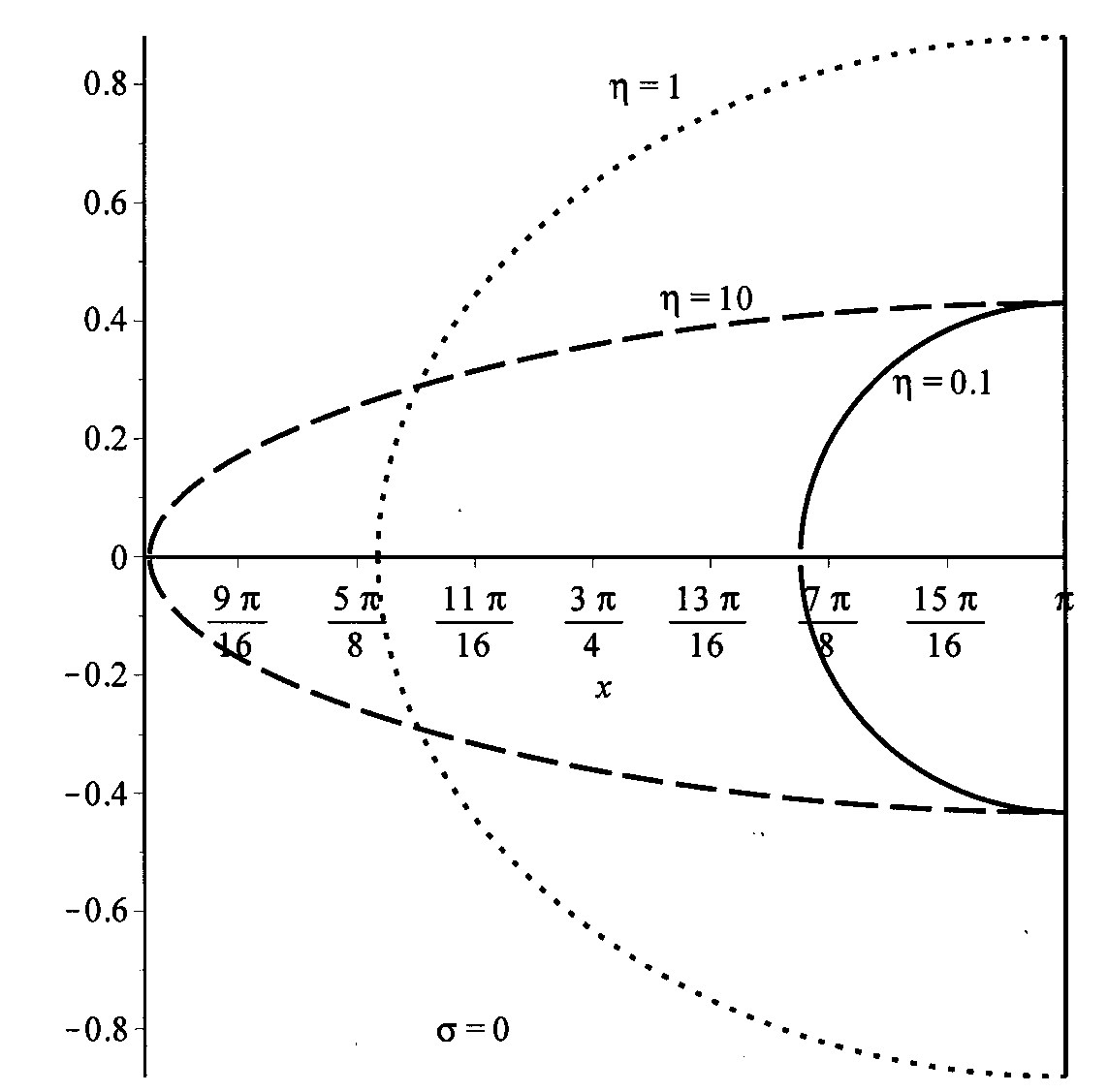}
\caption{the special case $\si=0$}
\end{figure}

\nin The empty interval $(0,x_*)$ is not represented. The  values of $x_*$ are respectively
\[x_*(\eta=0.1)=2.7\qq\qq x_*(\eta=1)=2 \qq\qq x_*(\eta=10)=1.6.\]
For $x=x_*$ the tangent is vertical since $P_x=0$ while for $x=\pi-$ it is horizontal since $P_x \to +\nf$.

\nin There remains the last case:

\begin{nth} For $\eta>0$ and $\si=-\cosh\tht\leq -1$ we have:
\beq \label{Eposi} 
\barr{llll}
(a) & \eta\in (0,e^{-\tht}) & \quad \cosh y=
\left\{\barr{ll}\dst \frac{\cos x-\eta}{\sqrt{\eta^2+2\,\si\,\eta+1}} & \quad x\in\,(0,x_-)
\\[4mm]\dst \frac{\eta-\cos x}{\sqrt{\eta^2+2\si\,\eta+1}}& \quad x\in\,(x_+,\pi)\earr\right.\\[12mm]
(b) & \eta\in(e^{-\tht},+\nf) 
& \dst e^{\eps y_{\eps}}=\frac{\eta-\cos x+\sqrt{2\eta(|\si|-\cos x)-\sin^2 x}}{\eta-1+\sqrt{2\eta(|\si|-1)}} \ \ x\in\,(0,\pi), & \\[8mm]\dst
(c) & \eta=e^{-\tht}=\cos x_* & \quad\hspace{7mm}  e^{\eps y_{\eps}}=\left\{\barr{ll}\dst \frac{\cos x-\cos x_*}{1-\cos x_*} & \qq x\in\,(0,x_*)\\[4mm]\dst \frac{\cos x_*-\cos x}{\cos x_*+1}& \qq x\in\,(x_*,\pi)\end{array}\right.\earr
\eeq
where $\eps=\pm 1$ and $x_{\pm}$ are defined by
\beq\label{xpm}
\cos x_{\pm}=\eta\mp\sqrt{\eta^2+2\si\,\eta+1}.\eeq
\end{nth}

\nin{\bf Proof:} From its derivative we see that $P_x^2$ decreases monotonically from $+\nf$ for $x\to 0+$ to $p_*^2=L^2\,e^{\tht}(\eta-e^{-\tht})$ for $x=x_*$, with $\cos x_*=e^{-\tht}$, and then it increases monotonically to $+\nf$ for $x\to \pi -$.

If $\eta\in (0,e^{-\tht})$ then the geodesic is defined 
for $x\in\,(0,x_-)\cup(x_+,\pi)$ with $x_-<x_*<x_+$ where $x_{\pm}$ are defined by (\ref{xpm}).

So if $x\in(0,x_-)$ we take for initial conditions $(x=x_-,\,y=0)$ which imply $S_+=S_-=L^2(\cos x_--\eta)$
and upon use of (\ref{alphai}) we get the first equation of (\ref{Eposi})(a). 

If $x\in(x_+,\pi)$ we take for initial conditions $(x=x_+,\,y=0)$ which imply $S_+=S_-=L^2(\cos x_+-\eta)$
and upon use of (\ref{alphai}) we get the second equation of (\ref{Eposi})(a).

The case $\eta=e^{-\tht}$ is quite special. In this case let us define $\cos x_*=e^{-\tht}$. The first integral gives
\[S_+=2L^2\,e^y\,(\cos x-\cos x_*) \qq\quad x\in (0,x_*)\]
and vanishes for $x\in\,(x_*,\pi)$. Taking for initial conditions $(x=0+,y=0)$ we get the first equation of  (\ref{Eposi})(b).

The second integral is
\[S_-=2L^2\,e^{-y}\,(\cos x-\cos x_*)\qq\quad x\in (x_*,\pi)\]
and vanishes for $x\in\,(x_*,\pi)$. Taking for initial conditions $(x=\pi-,y=0)$ we get the second equation of  (\ref{Eposi})(b).

The remaining case $\eta\in(e^{-\tht},+\nf)$ gives a geodesic defined for $x\in\,(0,\pi)$ and in view of the structure of the quadratic integrals we can take for initial conditions 
$(x=0+,y=0)$. The conservation of $S_-/L^2$ gives
\[e^{-y}(\eta-\cos x+\eps\sqrt{2\eta(|\si|-\cos x)-\sin^2 x})=\eta-1+\eps\sqrt{+2\eta(|\si|-1)}\]
and this concludes the Proof.$\quad\Box$

To conclude our analysis let us consider the case of a negative value for $\eta$ hence negative $E$. Since $P_x^2$ is invariant under the transformation $(E,\si,\xi)\to (-E,-\si,-\xi)$ it follows that $(-|E|,\si,\xi)$ is obtained from the above results for $(|E|,-\si,-\xi)$.

Let us give some examples of geodesic trajectories given by Propositions 5. For the case (a) we have

\begin{figure}[H]
\centering
\includegraphics[scale=0.5]{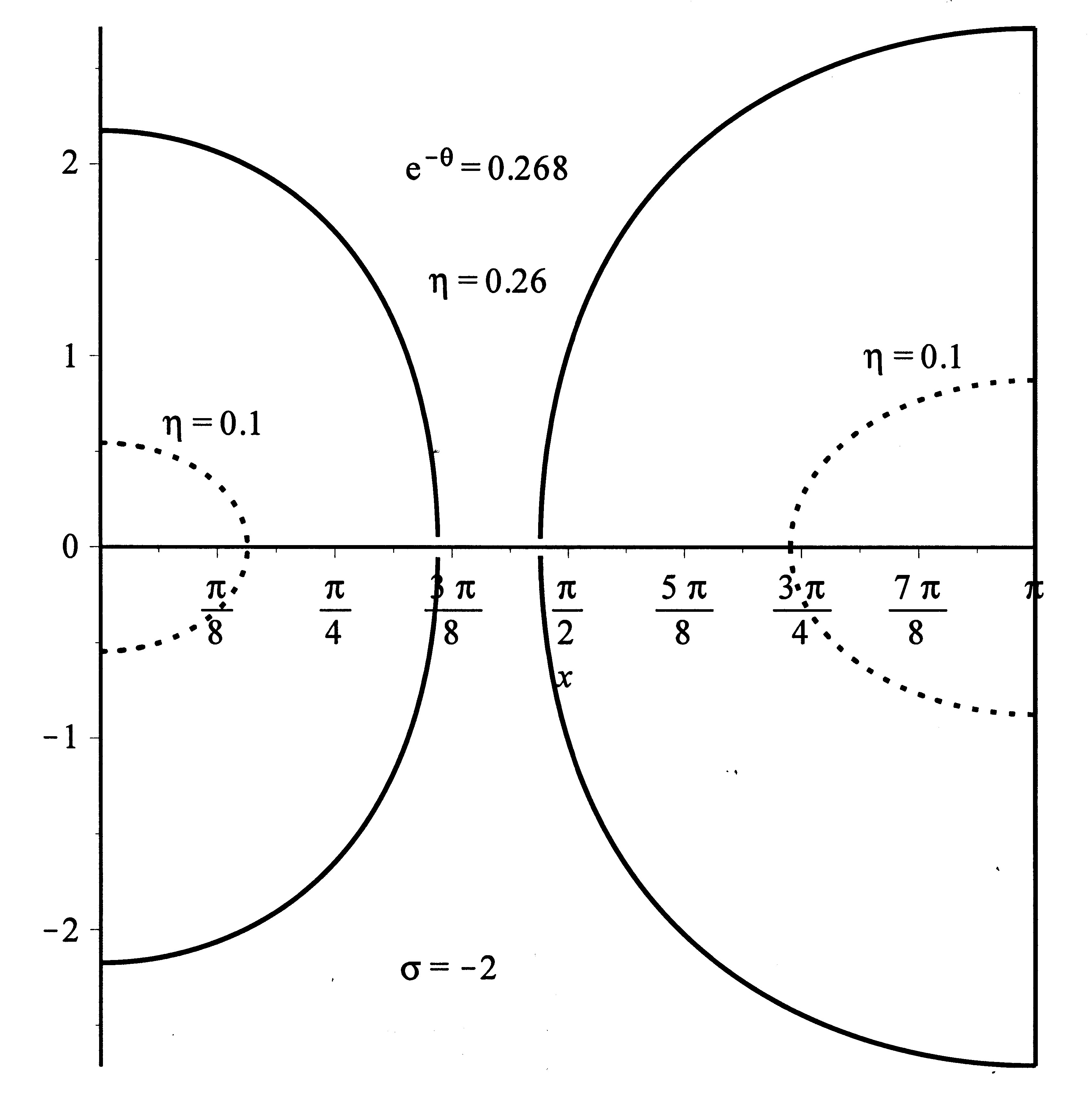}
\caption{the case $0<\eta< e^{-\tht}$}
\end{figure}

\nin{\bf Remarks:}
\brm
\item The $y$ coordinate is along the vertical.
\item The hamilton equations
\beq
\dot{x}=\frac{\sin^2 x}{1-\rho\,\cos x}\,P_x
\qq \dot{y}=\frac{\sin^2 x}{1-\rho\,\cos x}\,L\qq\Longrightarrow \qq\frac{dy}{dx}=\frac L{P_x}
\eeq
show that for $x\to 0+$ or $x\to\pi-$ the tangents to the geodesics are horizontal, while for $x=x_-$ or $x=x_++$ they are vertical.
\item The symmetry with respect to the axis $x=\pi/2$ which was apparent for vanishing energy has disappeared.
\erm

while for case (b) we have

\begin{figure}[H]
\centering
\includegraphics[scale=0.5]{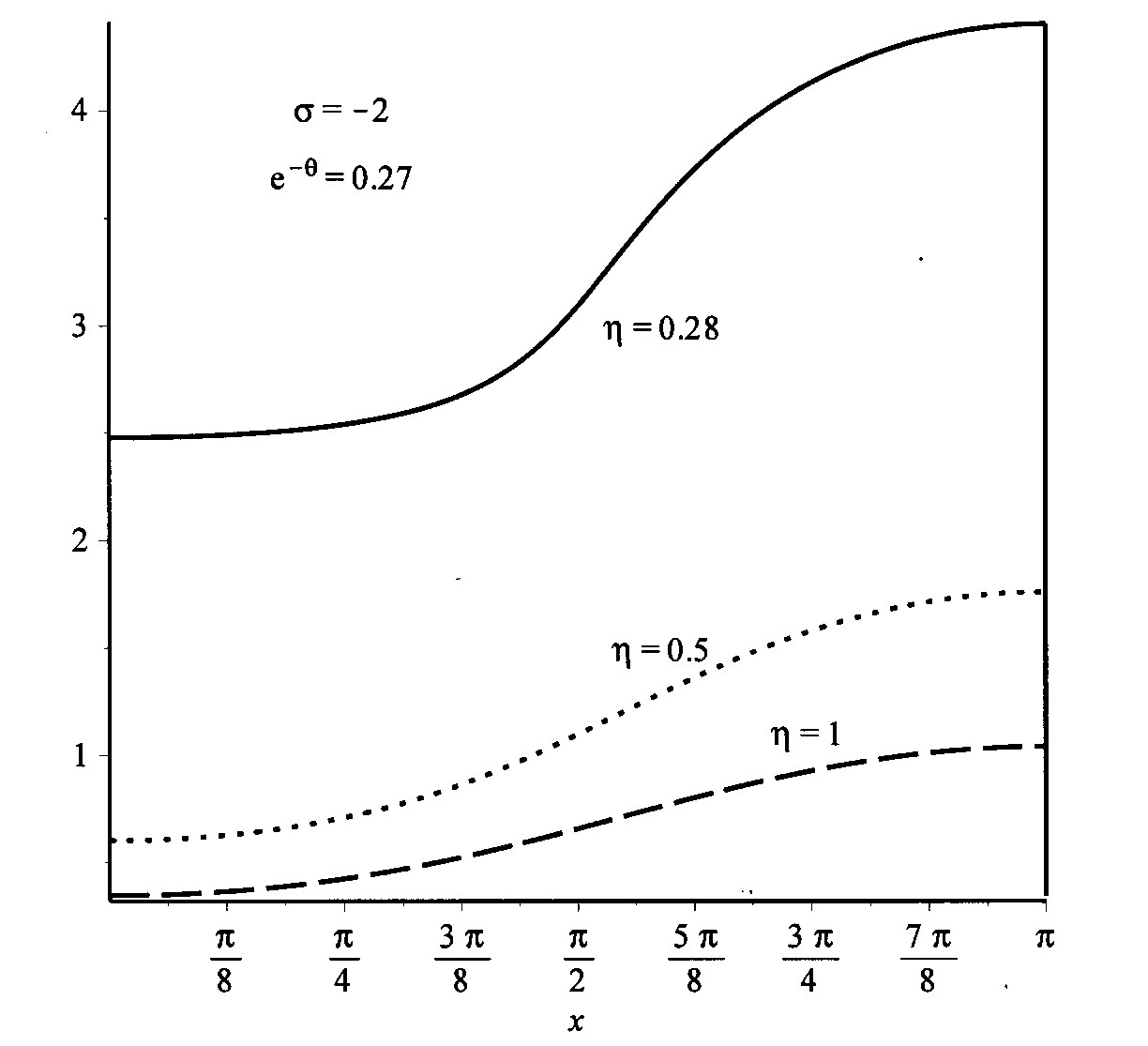}
\caption{the case $\eta>e^{-\tht}$}
\end{figure}
In this last drawing only the geodesics with positive velocity can be seen. The negative velocity ones are obtained from $y\to -y$.

For case (c), which is quite special, we have

\begin{figure}[H]
\centering
\includegraphics[scale=0.5]{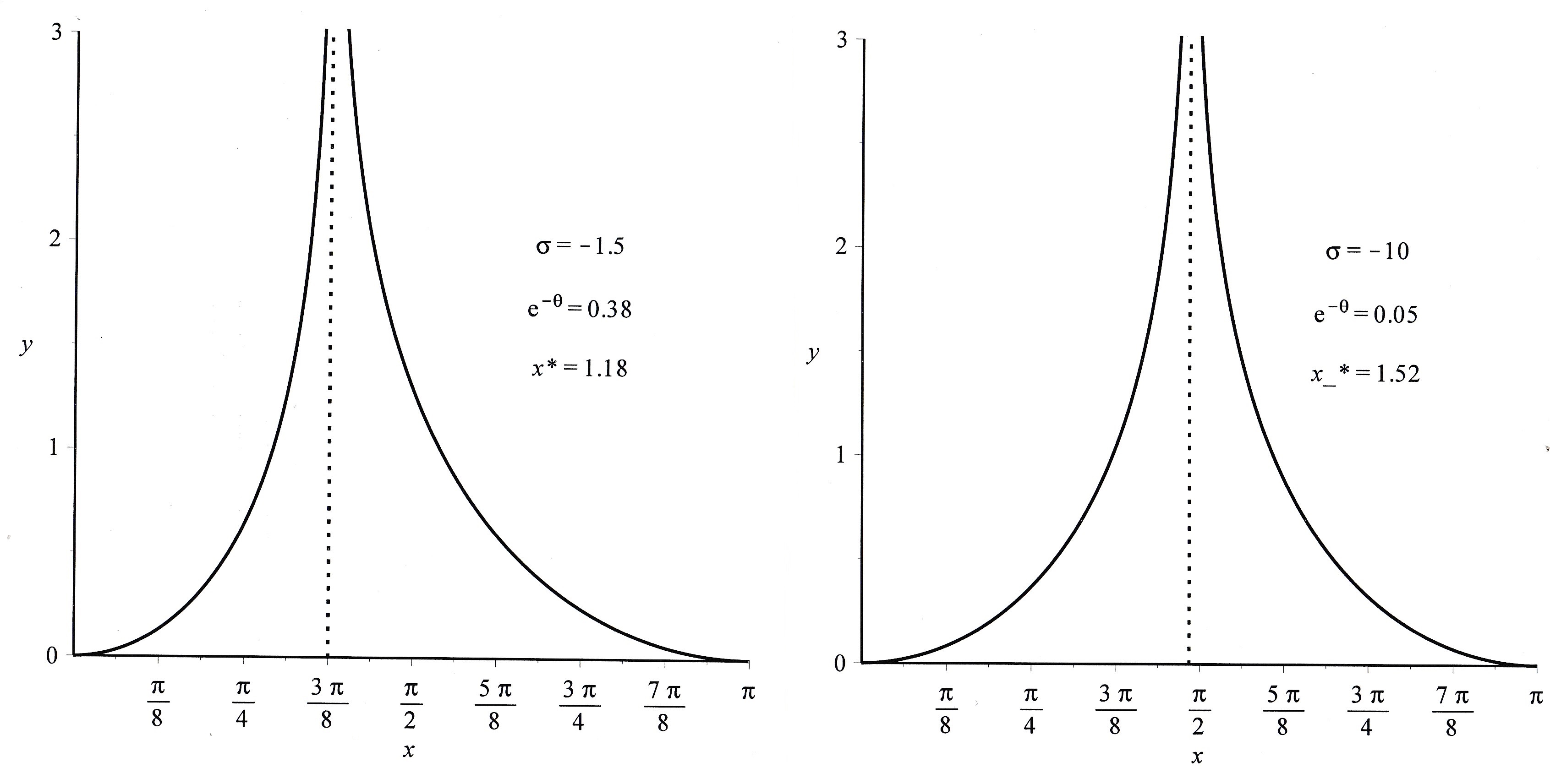}
\caption{the case $\eta=e^{-\tht}$}
\end{figure}
The geodesics are defined only on $x\in\,(0,\pi/2)\cup(\pi/2,\pi)$ and only the positve $y$ part of the graph is shown. Since the velocity vanishes for $x=x_*$ the corresponding line is some kind of a wall.

For the geodesics of vanishing energy (see Proposition 3) the main difference is that for the Figures 2 and 4 the line $x=\pi/2$ becomes an axis of symmetry.

\part{\bf\Large THE HYPERBOLIC CASE}

\setcounter{section}{0}

\section{Local structure}
Let us observe that for $A_0=0$ and $\mu=1$ the ODE $\,$
(\ref{odes})(ii) leads to three different cases:
\beq\label{odeHyp}
-A_1\,h\,h_x+A_2\,h_x=\frac{A_3}{2}(e^x+\eps\,e^{-x})
\eeq
where $\ \eps=0,\,\pm1.$ We have:
 
\begin{nth} The SI Koenigs systems
\beq
{\cal I}_1=\{H,\,P_y,\,S_1\}\qq\qq {\cal I}_2=\{H,\,P_y,\,S_2\}
\eeq
are given locally by three different metrics, For $\eps=0$ we have
\beq\label{metii0}
g_0=(e^{-x}+\rho\,e^{-2x})(dx^2+dy^2)
\eeq
with the integrals
\beq\label{intii0}
\left\{\barr{l}
S_1=+\cos y\,e^x\,P_x\,P_y+\sin y(e^x\,P_y^2-H)\\[4mm]
S_2=-\sin y\,e^x\,P_x\,P_y+\cos y(e^x\,P_y^2-H)
\earr\right.
\eeq
For $\eps=+1$ we have
\beq\label{metiiplus}
g_+=\frac{\cosh x+\rho}{\sinh^2 x}(dx^2+dy^2)
\eeq 
with the integrals
\beq\label{intiiplus} 
\left\{\barr{l}
S_1=+\cos y\,\sinh x\,P_x\,P_y+\sin y(\cosh x\,P_y^2-H)\\[4mm]
S_2=-\sin y\,\sinh x\,P_x\,P_y+\cos y(\cosh x\,P_y^2-H)
\earr\right.
\eeq
For $\eps=-1$ we have
\beq\label{metiimoins}
g_-=\frac{\sinh x+\rho}{\cosh^2 x}(dx^2+dy^2).
\eeq
with the integrals
\beq\label{intii3}
\left\{\barr{l}
S_1=+\cos y\,\cosh x\,P_x\,P_y+\sin y(\sinh x\,P_y^2-H)\\[4mm]
S_2=-\sin y\,\cosh x\,P_x\,P_y+\cos y(\sinh x\,P_y^2-H).\earr\right.
\eeq\end{nth}

\nin{\bf Proof:} The ODE (\ref{odeHyp}) is easily integrated to
\[-\frac{A_1}{2}\,h^2+A_2\,h=\frac{A_3}{2}(e^x-\eps\,e^{-x})+\wti{A}_4.\]
If $A_1$ vanishes the metric is of constant curvature, so we can take $A_1=-2$ and $A_2=-2 h_0$ ending up with
\[h-h_0=\pm \sqrt{\frac{A_3}{2}(e^x-\eps\,e^{-x})+A_4}\qq\Longrightarrow\qq h_x=\pm\frac{A_3}{4}\frac{e^x-\eps\,e^{-x}}{\sqrt{\frac{A_3}{2}(e^x-\eps\,e^{-x})+A_4}}.\]
So, up to an overall scaling, we get the three metrics given above and transforming the formulas of Matveev and Schevchishin \cite{ms} yields the quadratic integrals.$\quad\Box$

The metric $\rg_0$ corresponds to Koenigs type II metric
\[\Big(a\,e^{-w}+b\,e^{-2w}\Big)\,du\,dv\qq\qq w=\frac{u+v}{2}\]
when subjected to the coordinates change $(u=x+iy,\,v=x-iy)$ and an overall scaling.

The metric $\rg_-$ is still of type I when subjected to the coordinates change $(u=x+iy,\,v=-x+iy)$ up to scaling.

To recover the metric $\rg_+$ as a type I, up to scaling, we have to change the parameter $ a \to\,-ia$ and the coordinates $(u=x+i(y+\pi/2),\,v=-x+i(y-\pi/2))$.

Let us point out that the metric $\rg_0$ was first obtained in \cite{kk2} and the metric $\rg_+$ in \cite{kclm}. The ''new" metric $\rg_-$ is a close cousin of $\rg_+$ with the W-algebra:
\beq\barr{c}
\{P_y,S_1\}=S_2\qq \{P_y,S_2\}=-S_1\qq \{S_1,S_2\}=P_y\,(2P_y^2+2\rho\,H-\xi)\\[4mm] 
S_1^2+S_2^2=H^2-P_y^4-P_y^2(2\rho\,H-\xi).\earr
\eeq

In \cite{kk2} and \cite{kclm} it was shown that, keeping the same formulas for the quadratic integrals,  the following potentials could be added:
\beq
g_0:\quad V_0=\frac{\xi}{2(1+\rho\,e^{-x})}\qq 
g_+:\quad V_+=\frac{\xi}{2(\cosh x+\rho)}
\eeq
while for $g_-$ one easily obtains:
\beq
V_-=\frac{\xi}{2(\sinh(x)+\rho)}.
\eeq
Let us consider successively the three metrics obtained above including their potential.

\section{Global structure}

\subsection{The metric $\mathbf{g}_0$}

One has
\begin{nTh} The SI Koenigs systems
\beq
{\cal I}_1=\{H_0,\,P_y,\,S_1\}\qq\qq {\cal I}_2=\{H_0,\,P_y,\,S_2\}
\eeq
are globally defined on the manifold $M\cong {\mb H}^2$ 
 with the hamiltonian
 \beq\label{hamii1}
 H_0=\frac 1{2(1+\rho\,r^2)}\left(P_r^2+\frac{P_{\phi}^2}{r^2}+\xi\,r^2\right)
\eeq
and
\beq
(r,\phi)\in\,(0,+\nf)\times{\mb S}^1 \qq\qq (\rho,\,\xi)\in\,(0,+\nf)\times {\mb R}.\eeq
The integrals are
\beq\label{intii1}\left\{\barr{l}\dst 
S_1=+\cos(2\phi)\,P_r\,\frac{P_{\phi}}{r}+\sin(2\phi)\left(H_0-\frac{P_{\phi}^2}{r^2}\right)\\[5mm]\dst 
S_2=-\sin(2\phi)\,P_r\,\frac{P_{\phi}}{r}+\cos(2\phi)\left(H_0-\frac{P_{\phi}^2}{r^2}\right)\earr\right.
\eeq
\end{nTh}

\nin{\bf Proof:} Starting from the metric (\ref{metii0}) the change of coordinates
\[r=e^{-x/2}>0\qq\qq \frac y2=\phi\in\,{\mb S}^1\]
yields, up to scaling:
\[\rg=(1+\rho\, r^2)(dr^2+r^2\,d\phi^2)\qq (r,\phi)\in\,(0,+\nf)\times {\mb S}^1.\]
For this metric to be riemannian we must take $\,\rho\in\,(0,+\nf)$ leading to a conformal factor which is $C^{\nf}([0,+\nf))$ and to a negative scalar curvature 
\[R=-\frac{4\rho}{(1+\rho\,r^2)^3}.\]  
The integrals (\ref{intii0}) are easily deduced. 

To determine the manifold it is convenient to use cartesian coordinates
\[x_1=r\,\cos\phi\qq\qq x_2=r\,\sin\phi\]
which transform the metric into
\[\rg=\Big(1+\rho\,r^2\Big)\,\rg_0({\mb R}^2,{\rm can})\qq\qq g_0({\mb R}^2,{\rm can})=dx_1^2+dx_2^2.\]
Since the conformal factor is $C^{\nf}([0,+\nf))$ we conclude that the manifold is diffeomorphic to 
${\mb R}^2$.

Let us define 
\[P_1=\cos\phi\,P_r-\frac{\sin\phi}{r}\,P_{\phi}\qq 
P_2=\sin\phi\,P_r+\frac{\cos\phi}{r}\,P_{\phi}\qq L_3=x_1\,P_2-x_2\,P_1\]
which generate the $e(3)$ Lie algebra with
\[ \{P_1,P_2\}=0 \qq\qq \{L_3,P_1\}=-P_2 \qq\qq \{L_3,P_2\}=P_1.\]
In terms of these globally defined quantities in ${\mb R}^2$ we have
\beq\label{Hspe}
H=\frac 1{2(1+\rho(x_1^2+x_2^2))}\Big(P_1^2+P_2^2+\xi (x_1^2+x_2^2)\Big)\eeq
and for the integrals
\[\left(\barr{c} S_1\\[4mm] 2S_2\earr\right)=\left(\barr{c} P_1\,P_2\\[4mm] P_1^2-P_2^2\earr\right)+\frac{(-\rho(P_1^2+P_2^2)+\xi)}{(1+\rho(x_1^2+x_2^2))}\left(\barr{c} x_1\,x_2\\[4mm] x_1^2-x_2^2\earr\right)\]
concluding the proof.$\quad\Box$

\subsection{The metric $\mathbf{g}_+$}
\begin{nTh} The SI Koenigs systems
\[ {\cal I}_1=\{H,\,P_{\phi},\,S_1\}\qq\qq {\cal I}_2=\{H,\,P_{\phi},\,S_2\}\]
are globally defined on the manifold $M\cong {\mb H}^2$. The hamiltonian is
\beq\label{hamii2}
H=\frac 1{2(1+\rho\,\sinh^2\chi)}\left(\cosh^2\chi\,\pc^2+\frac{\pf^2}{\tanh^2\chi}+\xi\,\sinh^2\chi\right)
\eeq
with 
\[(\chi,\,\phi)\in\,(0,+\nf)\times {\mb S}^1 \qq\qq \rho\in\,(0,1)\cup (1,+\nf)\qq\qq \xi\in\,{\mb R}\] 
and the integrals
\beq\label{intii2}
\left\{\barr{l}\dst 
S_1=+\cos(2\phi)\,\pc\,\frac{\pf}{\tanh\chi}+\sin(2\phi)\left(H-\frac{(2-\tanh^2\chi)}{\tanh^2\chi}\pf^2\right)\\[4mm]\dst 
S_2=-\sin(2\phi)\,\pc\,\frac{\pf}{\tanh\chi}+\cos(2\phi)\left(H-\frac{(2-\tanh^2\chi)}{\tanh^2\chi}\pf^2\right).\earr\right.\eeq
\end{nTh}

\nin{\bf Proof:} In the metric (\ref{metiiplus}) we can take $x>0$ since the metric is even and we will change $\rho$ into $\wti{\rho}$. The scalar curvature is
\[R=-\wti{\rho}-\frac{(1-\wti{\rho}^2)}{(\cosh x+\wti{\rho})^3}(3\cosh^2 x+3\wti{\rho}\cosh x+ \wti{\rho}^2-1)\]
forbids $\wti{\rho}=1$ which would be of constant curvature. To get a riemannian metric we must therefore restrict $\wti{\rho}\in\,(-1,+\nf)\bs\{1\}$.

The change of coordinates
\[\chi=\ln\frac{1+\sqrt{u}}{1-\sqrt{u}}\in\,(0,+\nf)\qq \qq \frac y2=\phi\in\,{\mb S}^1\]
brings the metric (\ref{metiiplus}) to its final form
\[g=\frac{1+\rho\,\sinh^2\chi}{\cosh^2\chi}(d\chi^2+\sinh^2\chi\,d\phi^2) \qq\qq \rho=\frac{1+\wti{\rho}}{2}\in\,(0,+\nf)\bs\{1\}\]
The integrals in (\ref{intii2}) are obtained by transforming the formulas (\ref{intii0}). 

To study the global structure we need the canonical    embedding of ${\mb H}^2\subset{\mb R}^3$:
\[x_1=\sinh\chi\,\cos\phi \qq x_2=\sinh\chi\,\sin\phi  \qq x_3=\cosh\chi\qq \chi\in\,(0,+\nf) \quad \phi\in\,{\mb S}^1\]
and the globally defined objects
\[M_1=\sin\phi\,\pc+\frac{\cos\phi}{\tanh\chi}\,\pf\qq M_2=-\cos\phi\,\pc+\frac{\sin \phi}{\tanh\chi}\,\pf\qq 
M_3=\pf\]
which generate the $sl(2,{\mb R})$ Lie algebra
\[\{M_1,M_2\}=M_3\qq\qq \{M_2,M_3\}=-M_1\qq\qq \{M_3,M_1\}=-M_2.\]
One has
\[g_0({\rm H}^2,{\rm can})=dx_1^2+dx_2^2-dx_3^2=d\chi^2+\sinh^2\chi\,d\phi^2\]
so that our metric can be written
\[g=\frac{1+\rho\,\sinh^2\chi}{\cosh^2\chi}\,g_0({\rm H}^2,{\rm can})\]
and since the conformal factor is $C^{\nf}([0,+\nf))$ 
the manifold is diffeomorphic to ${\mb H}^2$.

The global definiteness on ${\mb H}^2$ follows from 
\[H_+=\frac 1{1+\rho(x_1^2+x_2^2)}\Big(x_3^2(M_1^2+M_2^2-M_3^2)+\xi(x_1^2+x_2^2)\Big)\]
while for the integrals we have
\[\left(\barr{c}S_1\\[4mm] 2S_2\earr\right)=
\left(\barr{c} -M_1\,M_2\\[4mm] -M_1^2+M_2^2\earr\right)+\frac{(1-\rho)[M_1^2+M_2^2+(x_2\,M_1-x_1\,M_2)^2]+\xi\,x_3^2}{x_3^2[1+\rho(x_1^2+x_2^2)]}
\left(\barr{c} x_1\,x_2\\[4mm] x_1^2-x_2^2\earr\right)\]
concluding the proof. $\quad\Box$

Let us conclude with the following remark: there is a singular limit relating $H_+$ and $H_0$ which is the following:
\beq
\chi=\mu\,r\qq P_{\chi}=\frac{P_r}{\mu} \qq \rho=\frac{\wti{\rho}}{\mu^2} \qq \xi=\frac{\wti{\xi}}{\mu^4}\qq\qq \mu \to 0+\eeq
and we have
\beq
\lim_{\mu\to 0+}\ \mu^2\,H_+(\chi,P_{\chi},\rho,\mu)=H_0(r,P_r,\wti{\rho},\wti{\xi}\,).\eeq
However, due to its singular nature, it is not useful for any proof.

\newpage Let us analyze the last case:

\subsection{The metric $\mathbf{g}_-$}
We have:

\begin{nth} The metric
\beq
\rg_-=\frac{\sinh x+\rho}{\cosh^2 x}(dx^2+dy^2)\eeq
is never defined on a manifold.
\end{nth}

\nin{\bf Proof:} Here we must take $x\in\,{\mb R}$. The metric, to be riemannian, requires $\sinh x+\rho>0$, but since the scalar curvature is
\[R=-\rho+\frac{(1+\rho^2)}{(\sinh x+\rho)^3}(3\sinh^2 x+3\rho\sinh x+\rho^2+1)\]
the end point $\sinh x+\rho=0$ will be a curvature singularity precluding any manifold.

This can be understood in a different way using the coordinates change
\[y=\phi\in\,{\mb S}^1 \qq x=\ln\tan\left(\frac{\tht}{2}\right): \qq x\in\,{\mb R}\,\to\,\tht\in\,(0,\pi)\] 
which transforms the metric into
\[\rg_-=\left(\rho-\frac 1{\tan\tht}\right)(d\tht^2+\sin^2\tht\,d\phi^2)=\left(\rho-\frac 1{\tan\tht}\right)\,\rg_0(S^2,{\rm can}).\]
We indeed get a metric conformal to the 2-sphere, but 
the conformal factor is singular at the geometrical poles  $\tht=0$ and $\tht=\pi$. $\quad\Box$

Let us determine the geodesic curves for the two 
complete metrics $\rg_0$ and $\rg_+$.

\section{Geodesics}

\subsection{The geodesics of $\mathbf{g}_0$}
Working with the hamiltonian (\ref{hamii1}) we have:

\begin{nth}\label{clo1} The geodesics are given by:
\beq\label{geodii0}
\barr{lcl} (a) \quad & E\geq \xi/2\rho \quad & 
\left\{\barr{cl} E=0 \quad & \dst \frac L{r^2}=\sqrt{|\xi|}\,\cos(2\phi)\\[4mm] E\neq 0 \quad & \dst \frac{L^2}{|E|\,r^2}=sign(E)+e\,\cos(2\phi)\earr\right.
\\[10mm] 
(b) \quad & E_+<E<\xi/2\rho \quad & \dst \frac{L^2}{E\,r^2}=1+e\,\cos(2\phi)\earr
\eeq
with 
\beq\label{formii}
e=\sqrt{\dst 1+\frac{L^2}{E^2}(2\rho\,E-\xi)} \qq\qq  
E_{\pm}=L^2(-\rho\,\pm \sqrt{\rho^2+\xi/L^2}).\eeq
Obviously for case (b) the geodesics are closed.
\end{nth}

\nin{\bf Proof:} From the hamiltonian (\ref{hamii1}) it follows that
\beq
P_r^2=2E+(2\rho\,E-\xi)r^2-\frac{L^2}{r^2}\qq (P_r^2)'=\frac 2{r^3}\Big((2\rho\,E-\xi)r^4+L^2\Big).
\eeq

For $\ 2\rho\,E\geq\xi\ $ the function $P_r^2$ is monotonically increasing from $-\nf$ to $+\nf$. It vanishes for 
\[r_*^2=\frac{L^2}{E+\sqrt{\De}} \qq\qq 
\De=E^2+L^2(2\rho\,E-\xi).\] 
Taking for initial conditions $(r=r_*,\,\phi=0)$ gives  $S_1=0$ and $S_2=-\sqrt{\De}$ and upon use of  (\ref{alphai}) we obtain
\[\frac{L^2}{r^2}=E+\sqrt{\De}\,\cos(2\phi).\] 
It follows that for $E=0$ and $E\neq 0$ we have obtained the equations in (\ref{geodii0})(a) which 
describe hyperbolas. 

For $2\rho\,E<\xi$ we must have $E>0$ and $\xi>0$. The derivative $(P_r^2)'$ has a simple zero for $\dst r_*^2=\frac L{\sqrt{\xi-2\rho\,E}}$. The function  $P_r^2$ increases from $-\nf$ to 
\[p_*^2=2(E-\sqrt{\xi-2\rho\,E})=2\frac{(E-E_-)(E-E_+)}{E+\sqrt{\xi-2\rho\,E}},\qq\qq E_-<0<E_+,\] 
where $\ E_{\pm}\ $ are defined in (\ref{formii}), and then decreases to $-\nf$. The sign of $p_*^2$ is therefore essential. 

If $\ E\in\,(0,E_+]\ $ we have $p_*^2< 0$ hence $P_r^2$ is always negative and there will be no geodesic. If $\ E\in\,[E_+,\,\xi/2\rho)\ $ we will have $p_*^2>0 $. The function $P_r^2$ will exhibit two simple zeroes $r_{\pm}$ such that $r_-<r_*<r_+$ and given by
\[r_{\pm}^2=\frac{E(1\pm e)}{\xi-2\rho\,E}\]
Taking for initial conditions $(r=r_-,\,\phi=0)$ and using (\ref{alphai}) we obtain
\[\frac{L^2}{r^2}=E+\sqrt{\De}\,\cos(2\phi)\]
from which we deduce (\ref{geodii0})(b). 
$\quad\Box$

\subsection{The geodesics of $\mathbf{g}_+$}
We have to study the positivity of  
\beq
\pc^2=2E+L^2+\si\,\tanh^2\chi-\frac{L^2}{\tanh^2\chi}\qq\chi>0 \qq \si=2(\rho-1)E-\xi,\eeq
while the geodesics, using (\ref{intii2}), are obtained from
\beq\label{alphaii}
S_1\,\sin(2\phi)+S_2\,\cos(2\phi)=E-\frac{(1+\cosh^2\chi)}{2\,\sinh^2\chi}\,\pf^2.
\eeq
Writing the energy conservation 
\beq
\cosh^2\chi\,\pc^2+\frac{L^2}{\tanh^2\chi}=2E\,\cosh^2\chi+\si\,\sinh^2\chi=2E+(2\rho\,E-\xi)\sinh^2\chi
\eeq
we obtain

\begin{lem}\label{lun} One has the following inequalities:
\beq
\si\leq 0\quad\Longrightarrow\quad E>0\qq\qq\qq \xi -2\rho\,E\geq 0 \quad\Longrightarrow\quad E>0.\eeq
\end{lem}
For the discussions to come it will be convenient to use, rather than $\chi$, the variable
\[u=\tanh^2\chi\ \in\ (0,1)\]
leading to
\beq
u\,\pc^2\equiv F(u)=\si\,u^2+2(E+L^2/2)u-L^2\qq\qq F'(u)=\si+\frac{L^2}{u^2}.\eeq

The discussion involves two cases, according to the sign of the parameter $\xi-2\rho\,E.$

\begin{nth} For $\xi-2\rho\,E\leq 0$ the geodesic equation 
\beq\label{geodii2}
\frac{L^2}{\tanh^2\chi}=E+\frac{L^2}{2}+\left|E+\frac{L^2}{2}\right|\,e\,\cos(2\phi)\qq\quad e=\sqrt{1+\frac{L^2\si}{(E+\frac{L^2}{2})^2}}\eeq
does not lead to a closed curve because $e>1$.
\end{nth}

\nin{\bf Proof:} Let us first consider the case $\si\geq 0$. Then the function $F$ increases monotonously from $-\nf$ to $-(\xi-2\rho\,E)$. So if $\xi-2\rho\,E\geq 0$ there is no geodesic, while for $\xi-2\rho\,E<0$ the function $F$ will be positive for $u\in(u_-,1)$ with
\beq\label{umoins}
u_-=\frac{L^2}{(E+L^2/2)+\sqrt{\De}}\qq\qq \De=(E+\frac{L^2}{2})^2+L^2\,\si.\eeq
The initial conditions $(u=u_-,\,\phi=0)$ give 
\[S_1=0\qq\qq S_2=E-L^2\left(\frac 1{u_-}-\frac 12\right)=-\sqrt{\De}\]
and upon use of (\ref{alphaii}) we get (\ref{geodii2}).

The next case is for $-L^2\leq \si <0$. Then $F'$ has a simple zero for $u_*=L/\sqrt{|\si|}\geq 1$. It follows that the variations of $F$ are the same as for $\si\geq 0.$ Since $\si<0$ we know that $E+L^2/2>0$ which allows to write (\ref{geodii2}) as 
\[\frac{L^2}{\tanh^2\chi}=\left(E+\frac{L^2}{2}\right)[1+e\,\cos(2\phi)].\]

The last case is for $\si<-L^2$. This time $F'$ has a simple zero for $u_*=L/\sqrt{|\si|}< 1$ so that F increases from $-\nf$ for $u \to 0+$ to $p_*^2$ for $u=u_*$ and then decreases to $-(\xi-2\rho\,E)\geq 0$, where $p_*^2=(L-\sqrt{|\si|})^2-\xi+2\rho\,E.$ It follows that $F$ exhibits one simple root $u_-$ (given by (\ref{umoins})) such that $0<u_-<u_*$. Imposing the initial conditions we get again (\ref{geodii2}).
$\quad\Box$

\vspace{2mm}
\nin{\bf Remarks:}
\brm
\item For $\si=0$ the geodesic equation does simplify into
\beq\label{geodii1}
\frac{L^2}{\tanh^2\chi}=(2E+L^2)\,\cos^2\phi \qq\quad E>0.\eeq
\item For $\si>0$ the energy may be negative, and for the special case where $E=-L^2/2$ the geodesic remains well defined since we have
\beq
\frac{L^2}{\tanh^2\chi}=\sqrt{\si}\,\cos(2\phi) \qq\quad \si=2(\rho-1)L^2-\xi>0.\eeq
\erm

The closed geodesics will appear now: 

\begin{nth}\label{quinze} If
\beq\label{posii2}
E\in\,\left[E_+,\frac{\xi}{2\rho}\right) \qq \& \qq   \xi-\rho\,L^2>0\eeq
where
\beq
E_+=L\Big[\sqrt{\xi+\rho(\rho-1)L^2}-(\rho-1/2)L\Big]
\eeq
the geodesic equation 
\beq\label{geodii5}
\frac{L^2}{\tanh^2\chi}=\left(E+\frac{L^2}{2}\right)\Big(1+e\,\cos(2\phi)\Big)\eeq
leads to a closed curve since $e$, still given by (\ref{geodii2}), is strictly smaller than one.
\end{nth}

\nin{\bf Proof:} The function $\pc^2$ for $u\to 0+$ starts from $-\nf$ and increases monotonously to $p_*^2=2E+L^2-2L\sqrt{-\si}$ for $u=u_*<1$ and then decreases monotonously to $-(\xi-2\rho\,E)$ for $u\to 1-$. This time let us consider the case where $\xi-2\rho\,E>0.$ If $p_*^2<0$ no geodesic is allowed, hence let us take $p_*^2\geq 0$. It follows that $\pc^2$ will be positive for $u\in\,(u_-,\,u_+)$ such that $0<u_-<u_*<u_+<1$ with
\[u_-=\frac{L^2}{E+\frac{L^2}{2}+\sqrt{\De}}\qq\qq\qq  u_+=\frac{L^2}{E+\frac{L^2}{2}-\sqrt{\De}}.\]
Taking for initial conditions $(u=u_-,\,\phi=0)$ gives
\[S_1=0\qq\qq S_2=E-L^2\left(\frac 1{u_-}-\frac 12\right)=-\sqrt{\De}\]
and we conclude using (\ref{alphaii}).

One has to discuss the initial algebraic conditions:
\[\xi-2\rho\,E>0 \qq\qq \si<-L^2 \qq\qq p_*^2=(E+L^2/2)^2+L^2\,\si\geq 0\]
to show that they lead to (\ref{posii2}). The analysis  involves elementary algebra and will be skipped.
$\quad\Box$

\vspace{3mm}
\nin{\bf Remark:} Let us observe that the geodesics of this metric $\rg_+$ were discussed in \cite{kclm}. These authors write the  metric
\[g=\frac{2\cosh(2x)+b}{\sinh^2(2x)}(dx^2+dy^2)\qq x>0 \qq  2y\in{\mb S}^1 \qq  b>-2\]
which is nothing but our metric $\rg_+$ given by (\ref{metiiplus}). In order to describe the geodesics they change the coordinates $(x,y)$ into $(r,\,\tht)$ given by \footnote{Correcting an obvious typo.}
\[r=\frac{\sqrt{2\cosh(2x)+b}}{2\sinh(2x)}\in\,(0,+\nf) \qq\qq \tht=2y\in {\mb S}^1.\]
However, since we have
\[\frac{dr}{dx}=-\frac{(1+2br^2)}{\sinh^2(2x)\,\sqrt{2\cosh(2x)+b}},\]
we realize that for $b\in\,(-2,0)$ this is not a local diffeomorphism hence $r$ is not a coordinate, at variance with our choice of coordinates which is valid for $b>-2$.
 Of course for $b>0$ we are in complete agreement with \cite{kclm} albeit our coordinate $\chi$ is somewhat different from their coordinate $r$ while our $\phi$ and their $\tht$ are the same.

\part{\bf\Large THE AFFINE CASE}

\setcounter{section}{0} 

\section{Local structure}
The local structure, already found in \cite{kk1} and \cite{kk2}, is given by

\begin{nth} The SI Koenigs systems
\beq
{\cal I}_1=\{H,\ P_y,\ S_1\}\qq\qq {\cal I}_2=\{H,\ P_y,\ S_2\}\eeq
are given locally by
\beq\label{kkiii}
H=\frac{(a_2 x+a_1)^2}{a_2x^2+2a_1x+a_0}(P_x^2+P_y^2)\eeq
with
\beq\label{intkkiii}\left\{\barr{l}
S_1=(a_2 x+a_1)\,P_x\,P_y-y(H-a_2\,P_y^2) \\[4mm] 
2S_2=(a_2 x^2+2a_1 x)P_y^2+2y(a_2 x+a_1)\,P_x\,P_y-y^2(H-a_2\,P_y^2).\earr\right.
\eeq
\end{nth}

\nin{\bf Proof:} The ODE (\ref{odes}) (iii) for $A_0=0$ becomes
\[-A_1 h h_x+A_2\,h_x=A_3 x+A_4\qq\Longrightarrow\qq -\frac{A_1}{2}\,h^2+A_2\,h=\frac{A_3}{2}\,x^2+A_4 x+A_5.\]
Since $A_1$ cannot vanish we set $A_1=-2$ and $A_2=-2h_0$ which leads us to
\[h=h_0\pm\sqrt{a_2x^2+2a_1x+a_0}\qq\qq h_x=\pm\frac{a_2 x+a_1}{\sqrt{a_2x^2+2a_1x+a_0}}\]
and to the metric 
\[\rg=P(x)\ \frac{dx^2+dy^2}{(a_2 x+a_1)^2}\qq\qq P(x)=a_2 x^2+2a_1 x+a_0\]
which implies the hamiltonian (\ref{kkiii}).$\quad\Box$

Let us compare with Koenigs results. His type III  metric subjected to the coordinates change $\,(u=x+iy,\,v=-x+iy)$ gives
\beq
\rg_K=\left(\frac a{(u-v)^2} +b\right)\,du\,dv \qq\Longrightarrow \qq \rg_K=\left(\frac a{4x^2}+b\right)(dx^2+dy^2)\eeq
while the change of coordinate $a_2\,x+a_1\,\to\,x$, possible for $a_2\neq 0$, transforms our metric into:
\beq
\rg=\left(1+\frac{a_0-a_1^2}{x^2}\right)\,(dx^2+dy^2).
\eeq
Both agree (up to an overall scaling) for $b\neq 0$ while the case $b=0$ must be excluded since one recovers a constant curvature metric.

Koenigs type IV metric, up to the same coordinates change as above gives     
\beq
\rg_K=(u+v)\,du\,dv\qq\Longrightarrow\qq \rg_K=2x(dx^2+dy^2).\eeq

This should be compared with our metric for $a_2=0$. Then we must have $a_1\neq 0$, otherwise the metric becomes flat, and the change of coordinate $x+a_0/2 a_1\,\to\, x$ gives
\[g=x(dx^2+dy^2)\]
which is Koenigs type I as pointed out in \cite{kk1}. Therefore the affine case unifies at the same time Koenigs types III and IV.

\section{Global structure}
The scalar curvature
\[\frac R2=\frac{\De}{P^3}\Big(3\,(a_2 x+a_1)^2-\De\Big)\qq\qq  \De=a_1^2-a_0\,a_2\neq 0\]
shows:
\brm
\item That to avoid a flat metric we must impose $\De\neq 0$.
\item That a simple zero of $P$ is a curvature singularity.
\erm

The global structure follows from

\begin{nTh} The SI Koenigs systems
\beq
{\cal I}_1=\{H,\ P_y,\ S_1\}\qq\qq {\cal I}_2=\{H,\ P_y,\ S_2\}\eeq
are globally defined on the manifold $M\cong {\mb H}^2$. The hamiltonian is
\beq\label{Hiii}
H=\frac 1{2(1+\rho\,u^2)}\Big( u^2(P_u^2+P_y^2)+\xi\Big)\qq (u,y)\in(0,+\nf)\times{\mb R}
\qq \rho\in(0,\nf)\eeq
and the integrals
\beq\label{intiii}\left\{\barr{l} S_1=u\,P_u\,P_y-y(2\rho\,H-P_y^2)\\[4mm]
2S_2=-u^2\,P_y^2+2yu\,P_u\,P_y-y^2(2\rho\,H-P_y^2).\earr\right.
\eeq 
We have the algebraic relations
\beq\label{algiii1}
\{P_y,S_2\}=S_1\qq \{P_y,S_1\}=P_y^2-2\rho\,H\qq \{S_1,S_2\}=(2S_2+2\,H-\xi)P_y\eeq
and
\beq\label{algiii2}
S_1^2+2(2\rho\,H-P_y^2)S_2=(2\,H-\xi)P_y^2.\eeq
\end{nTh}

\nin{\bf Proof:} Let us organize the discussion according to the values of $a_2$. 

If $a_2=0$ then $a_1\neq 0$ (otherwise the metric is flat) so let us take $a_1=1$. The coordinate 
$u=x+a_0/2$ gives the type I metric $g=u(du^2+dy^2)$. This metric is riemannian iff  $u>0$ and $y\in{\mb R}$. Its scalar curvature being $R=u^{-3}$ it follows that 
the end-point $u=0$ is a curvature singularity precluding any manifold.

If $a_2=1$ defining $u=x+a_1$ gives for the type II metric
\[g=(u^2-\De)\frac{du^2+dy^2}{u^2}\qq \De=a_1^2-a_0\qq  u>0\qq y\in{\mb R}.\] 

Using the embedding ${\mb H}^2\subset {\mb R}^3$ given in \cite{vds}:
\[x_1=\frac yu\qq x_2=\frac 1{2u}(u^2+y^2-1)\qq x_3=\frac 1{2u}(u^2+y^2+1)\qq u>0 \quad y\in{\mb R}\]
leads to
\[g_0({\mb H}^2)=dx_1^2+dx_2^2-dx_3^2=\frac{du^2+dy^2}{u^2}\qq\Longrightarrow\qq g=(u^2-\De)\,g_0({\mb H}^2).\]
So if $\De>0$ the conformal factor vanishes for $u=\sqrt{\De}$ implying a curvature singularity while if $\De<0$ the conformal factor never vanishes and the manifold 
is diffeomorphic to ${\mb H}^2$. Defining $\rho=-1/\De$, up to a scaling, we get the metric (\ref{Hiii}).

If $a_2=-1$ defining $u=x-a_1$ gives for the metric
\[g=(\De-u^2)\,g_0({\mb H}^2)\qq\qq \De=a_1^2+a_0.\]
If $\De>0$ the end-point $u=\sqrt{\De}$ will be singular, while for $\De<0$ we must change the overall sign to be riemannian and we are back to the case $a_2=1$.

The integrals are easily transformed from (\ref{intkkiii}) and give (\ref{intiii}). They allow again for a potential, which does not modify their structure. The relations (\ref{algiii1}) and (\ref{algiii2}) are then easily checked.

The global structure is best displayed using the generators defined in \cite{vds}:
\[\barr{l}
M_1=u\,p_u+y\,P_y\\[4mm]\dst 
M_2=uy\,P_u+\frac{(y^2-u^2-1)}{2}\,P_y\\[4mm]\dst 
M_3=uy\,P_u+\frac{(y^2-u^2+1)}{2}\,P_y\earr\]
which generate the $sl(2,{\mb R})$ Lie algebra. The relations
\[H=\frac 1{2(1+\rho\,u^2)}(M_1^2+M_2^2-M_3^2+\xi)\qq\qq u=\frac{x_2+x_3}{1+x_1^2}\]
and
\[S_1=-M_1(M_2-M_3)-2\rho\,y\,H\qq 2S_2=M_2^2-M_3^2-2\rho\,y^2\,H\qq 
y=\frac{x_1(x_2+x_3)}{1+x_1^2}\]
show that this system is globally defined on ${\mb H}^2$. $\quad\Box$

\section{Geodesics}
From the hamiltonian (\ref{Hiii}) we get
\beq
P_u^2=\frac{2E-\xi}{u^2}+2\rho\,E-L^2\qq\qq E\in{\mb R}\qq L>0,
\eeq
while the integrals are
\beq\label{Siii}
S_1=L\,uP_u+y(L^2-2\rho\,E)\qq\qq 2S_2=-L^2\,u^2+2Ly\,uP_u+y^2(L^2-2\rho\,E).
\eeq
We have for first case
\begin{nth}  If $\ 2E<\xi\ $  and $\ 2\rho\,E>L^2\ $  the geodesic equation is
\beq\label{G1iii}
u^2-\frac{(2\rho\,E-L^2)}{L^2}\,(y-y_0)^2=u_*^2 \qq\qq u\in\,(u_*,\,+\nf)\eeq
where
\[u_*=\sqrt{\frac{\xi-2\,E}{2\rho\,E-L^2}}.\]
\end{nth}

\nin{\bf Proof:}  For $2E<\xi$ the classical motion is possible iff $2\rho\,E-L^2>0$ and for $u\in\,(u_*,+\nf).$ 
Taking for initial conditions $(u=u_*,\,\ y=y_0)$ the conservation of $S_1$ gives
\[S_1=-y_0(2\rho\,E-L^2)=L\,uP_u-y(2\rho\,E-L^2)\]
which implies (\ref{G1iii}). $\quad\Box$

We have for the second case
\begin{nth}
If $\ 2E=\xi\ $ the geodesic degenerates into the lines
\beq\label{G2iii}
u=\sqrt{\frac{(2\rho\,E-L^2)}{L^2}}\, |y-y_0| \qq\qq u\in\,(0,\,+\nf).\eeq
\end{nth}

\nin{\bf Proof:} Using the Hamilton equations
\[\dot{u}=\pm\sqrt{(2\rho\,E-L^2)}\frac{u^2}{1+\rho\,u^2}\qq\qq \dot{y}=\frac{Lu^2}{1+\rho\,u^2}\]
we get
\[\frac{du}{dy}=\pm\,\sqrt{\frac{(2\rho\,E-L^2)}{L^2}}\]
which implies (\ref{G2iii}). These are the asymptotes of the hyperbolas (\ref{G1iii}). $\quad\Box$

Let us conclude with the last case:

\begin{nth} If $\ 2E>\xi\ $ we have three possible types of geodesics:
\beq\barr{cll} 
2\rho\,E>L^2 & \qq\dst u^2+u_*^2=\frac{(2\rho\,E-L^2)}{L^2}\,(y-y_0)^2 & \qq u\in\,(0,+\nf)\\[5mm] 
2\rho\,E=L^2 & \qq\dst |y-y_0|=\frac L{2\sqrt{2E-\xi}}\,u^2  & \qq u\in\,(0,+\nf)\\[5mm] 
2\rho\,E<L^2 & \qq\dst u^2+\frac{(2\rho\,E-L^2)}{L^2}\,(y-y_0)^2=u_*^2 & \qq u\in\,(u_*,\,+\nf)\earr
\eeq
where
\[u_*=\sqrt{\frac{2E-\xi}{|2\rho\,E-L^2|}}.\]
\end{nth}

\nin{\bf Proof:} In the first case the positivity of $P_u^2$ allows for $u\in(0,+\nf)$. Taking for initial 
conditions $(u=u_*,\,y=y_0)$ and using as in the proof of Proposition 6 the conservation of $S_1$ we 
get the first geodesic equation. 

In the second case, resorting to Hamilton equations we get
\[\frac{dy}{du}=\pm\,\frac L{\sqrt{2E-\xi}}\,u.\]

In the last case the positivity of $P_u^2$  requires $u\in\,(u_*,\,+\nf)$. Taking the same initial conditions 
as above one gets the required result.$\quad\Box$

\vspace{2mm}
\nin{\bf Remarks:}
\brm
\item In all the cases above we have checked that the conservation of $S_2$ gives the same result 
as the conservation of $S_1$.
\item All the conics appear for the geodesic equations obtained here, particularly circles. This can be 
compared with the geodesics of the hyperbolic plane which are either circles or lines $(u>0,\,y=y_0)$.
\erm

\part{\bf \Large QUANTUM ASPECTS}

\setcounter{section}{0}
\section{Carter quantization}
We can go a step further and examine the quantization of SI models. We will adhere to the simplest concept of ``quantum superintegrability" which is the following: at the classical level we have seen that the relations
\beq\label{SIclass}
\{H,P_y\}=0\qq\qq \{H,S_1\}=0 \qq\qq \{H,S_2\}=0
\eeq
do hold. Quantizing means that to the previous classical observables we associate, by some recipee, operators  
$\wh{H},\ \wh{P}_y,\ \wh{S}_1,\ \wh{S}_2$ acting in the Hilbert space built up on the corresponding curved manifold.

The system will be defined as quantum superintegrable iff
\beq\label{SIquant}
[\wh{H},\wh{P}_y]=0\qq\qq [\wh{H},\wh{S}_1]=0 \qq\qq [\wh{H},\wh{S}_2]=0.\eeq
While the relations (\ref{SIclass}) are rigorous, the relations (\ref{SIquant}) are most often checked only formally, which is of course required, but hides the delicacies involved in a proper definition of their self-adjoint extensions.  

The simplest and most natural quantization is certainly Carter's (or minimal) quantization (see \cite{dv}). Denoting by a hat the quantum operators and setting $\hbar=1$, the quantization rules are:
\beq\label{Carter}
\wh{Q^i\,P_i}=-\frac i2(Q^i\circ\nabla_i+\nabla_i\circ Q^i)\qq \wh{S^{ij}\,P_i\,P_j}=-\nabla_i\circ S^{ij}\circ\nabla_j.
\eeq 
As a consequence we have:

\begin{nth} All of the classical SI Koenigs systems remain {\em formally} SI at the quantum level using Carter quantization.
\end{nth}

\nin{\bf Proof:} As shown in \cite{dv} in equation (3.8), since $P_y$ is generated by a Killing vector, we have
\beq\label{conserv1}
[\wh{H},\wh{P_y}]=0.
\eeq
For the quadratic observables, as shown in \cite{Ca},  
if $S$ is a quadratic Killing-Stackel tensor one has
\[[\wh{H},\wh{S}]=\frac 23\Big((\na_i\,B^{ij})\circ \na_j+\na_j\circ (\na_iB^{ij})\Big)\] 
where
\[B^{ij}=S^{k[i}\,{\rm Ric}_{\,kl}\,g^{j]l}.\] 
For a two dimensional metric which is diagonal, as it is the case for all of the Koenigs metrics, the Ricci tensor is always diagonal. It follows that the tensor $B$ vanishes identically. Therefore the classical conservation laws for $S_1$ and $S_2$ are lifted up to the quantum conservation laws
\beq\label{conserv2}
[\wh{H},\wh{S_1}]=0\qq\qq [\wh{H},\wh{S_2}]=0\eeq
and this concludes the proof. $\quad\Box$

\vspace{2mm}
{\bf Remarks:} 
\brm
\item Let us put some emphasis on the formal character of the proof. Indeed we are working with unbounded operators defined only on dense subspaces of the Hilbert space. Computing their commutators {\em non-formally} is a very difficult task.
\item One could use, as an alternative quantization, the so-called conformally equivariant quantization \cite{do}. Then (\ref{conserv1}) is still valid while relations (\ref{conserv2}) no longer hold for this quantization.
\erm

Before diving into the hamiltonian spectrum it is of some interest to consider the action coordinates which are of conceptual interest.

\section{Action coordinates for $\mathbf{g}_0$}

We have

\begin{nth} The action coordinates, for the closed geodesics obtained in Proposition \ref{clo1}, are given by
\beq\label{aav}
I_{\phi}=L\qq J\equiv I_r+I_{\phi}=\frac E{\sqrt{\xi-2\rho\,E}},\eeq
and the hamiltonian is
\beq\label{aaH}
H(J)=J\Big(\sqrt{\xi+\rho^2\,J^2}-\rho\,J\Big)\eeq
while the quadratic integrals are
\beq\label{aaS}
S_1=0\qq\qq S_2=-\sqrt{J^2-I_{\phi}^2}\Big(\sqrt{\xi+\rho^2\,J^2}-\rho\,J\Big).\eeq
\end{nth}

\nin{\bf Proof:} The Hamilton-Jacobi equation, starting from the action 
\[S=W(r)+L\,\phi-E\,t,\]
gives trivially $I_{\phi}=L$. It remains to compute
\[I_r=\frac 1{2\pi}\oint W'\,dr=\frac 2{\pi}\int_{r_-}^{r_+}\ W'\,dr \qq\qq 
W'=\sqrt{2(1+\rho\,r^2)E-\xi\,r^2-\frac{L^2}{r^2}}.\]
The first change of variable 
\[r \to \tht: \qq\frac{L^2}{E\,r^2}=1+e\,\cos\tht 
\qq\Longrightarrow\qq I_r=\frac{L\,e^2}{\pi}\int_0^{\pi}\frac{\sin^2\tht}{(1+e\,\cos\tht)^2}\,d\tht,\]
and the second change $\dst t=\tan\frac{\tht}{2}$ 
gives eventually
\[I_r=\frac{4Le^2}{\pi}\int_{-\nf}^{+\nf}\frac{t^2}{(1+t^2)[1+e+(1-e)t^2]^2}\,dt\]
which is computed using the residue theorem and gives (\ref{aav}). As we have seen in Proposition \ref{clo1} we have $\,E\in\,[E_+,\xi/2\rho)$ where
\[E_+=L^2(-\rho\, + \sqrt{\rho^2+\xi/L^2}).\]
Differentiating  
\[J=\frac E{\sqrt{\xi-2\rho\,E}}\qq\qq J=I_r+I_{\phi}\geq L\]
shows that $J(E)$ is a strictly increasing bijection from $E\in[E_+,\xi/2\rho)$ to $J\in [L,+\nf)$. The inversion needed for $E(J)$ is elementary and gives (\ref{aaH}). 

The integrals follow from the initial conditions which had given $S_1=0$ and $S_2=-\sqrt{\De}$. Expressing $\,S_2$ in terms of the action variables gives (\ref{aaS}).  $\quad\Box$

\nin{\bf Remarks:}
\brm
\item The hamiltonian is degenerate, a typical feature of SI systems.
\item The closed geodesics stem from the potential: indeed, if $\xi=0$ there are no ellipses at all and since we have $\xi>0$ the radial component of the force derived from the potential is attractive and given by
\[F_r=-\frac{\xi\,r}{(1+\rho\,r^2)^2}.\]
\item The knowledge of the action-angle coordinates establishes its bi-hamiltonian structure as shown by 
Bogoyavlenskij \cite{Bo}. 
\erm

Let us determine, for the classical hamiltonian $H_0$  given by (\ref{hamii1}), the discrete spectrum of its quantum extension.

\section{Point spectrum for the hamiltonian on $\mathbf{g}_0$} 
Using Carter quantization we have 
\beq
\wh{H_0}=-\frac 12\,\nabla_i\circ g^{ij}\circ\nabla_j+V(r)=-\frac 12\,\De+V(r)\qq\quad V(r)=\frac{\xi\,r^2}{2(1+\rho\,r^2)}.
\eeq

\begin{nth} The point spectrum of $\wh{H_0}$ is given by
\beq\label{spectrum0}
E_{n,m}=\wti{J}\Big(\sqrt{\xi+\rho^2\,\wti{J}^2}-\rho\,\wti{J}\Big) \qq \wti{J}=2n+|m|+1\qq (n,m)\in {\mb N}\times{\mb Z},\eeq
and the eigenfunctions 
\beq\label{eigfcts0}
\Psi_{n,m}(r,\phi)=e^{-\ze/2}\,\ze^{|m|/2}\,L_n^{|m|}(\ze)\,e^{im\phi}\qq\qq \ze=\sqrt{\xi-2\rho\,E}\ r^2  
\eeq
are expressed in terms of Laguerre polynomials.
\end{nth}

\nin{\bf Proof:} We have to solve the eigenvalue problem
\[(\wh{H_0}-E)\,\Psi(r,\phi)=-\frac 1{2(1+\rho\,r^2)}\left(\pt^2_r+\frac 1r\,\pt_r+\frac 1{r^2}\,\pt^2_{\phi}\right)\Psi(r,\phi)+(V(r)-E)\,\Psi(r,\phi)=0\]
for which we can take
\[\,\Psi(r,\phi)=e^{im\phi}\,\psi(r),\quad m\in{\mb Z} \quad\Longrightarrow\quad \wh{\pf}\,\,\Psi(r,\phi)=m\,\,\Psi(r,\phi)\]
The resulting radial ODE 
\[\left(\pt^2_r+\frac 1r\,\pt_r-\frac{m^2}{r^2}+2E-(\xi-2\rho\,E)\,r^2\right)\psi(r)=0,\]
upon the changes
\[\ze=\sqrt{\xi-2\rho\,E}\ r^2\qq\qq \psi(r)=e^{-\ze/2}\ze^{|m|/2}\,R(\ze)\]
gives for $R$ the confluent hypergeometric ODE
\[\ze\,R''+(c-\ze)R'-aR=0 \qq\qq a=\frac 12\left(|m|+1-\frac{E}{\sqrt{\xi-2\rho\,E}}\right) \qq c=|m|+1.\]
Its two independent solutions are denoted in \cite{be} as
$\Phi(a,c;\ze)$ and $\Psi(a,c;\ze)$ and we have to impose that the eigenfunctions are square summable i. e.
\[\int_0^{+\nf}\,(1+\rho\,\ze)\,\ze^{|m|}\,|R(\ze)|^2<+\nf.\] 

Taking into account the 

The general solution, square integrable for $\ze\to 0+$, 
is
\[\left\{\barr{ll} m=0  & \quad R(\ze)=A_0\,\Phi(a_0,1;\ze)+B_0\,\Psi(a_0,1;\ze)\\[4mm] m\neq 0 & \quad R(\ze)=A_m\,\Phi(a,|m|+1;\ze)\earr\right.\]
For $\ze \to +\nf$ we have
\[\Phi(a,c;\ze)=\frac{\G(c)}{\G(a)}\,\ze^{a-c}\,e^{\ze}\left[1+{\cal O}\left(\frac 1{\ze}\right)\right]
\]
and the exponential increase destroys the square summability. This can be avoided iff the parameter $a=-n$ with $n\in{\mb N}$ since then $\Phi$ reduces to a polynomial. 

This gives \footnote{The similarity of this quantum relation with the classical relation (\ref{aav}) is really striking.}
\[\wti{J}=\frac E{\sqrt{\xi-2\rho\,E}}=2n+|m|+1\qq\qq n\in{\mb N} \qq m\in{\mb Z}.\]
Squaring produces a second degree equation for $E$ giving the expected spectrum (\ref{spectrum0}). 

The relations 
\[\Phi(-n;|m|+1;\ze)={n+|m| \choose n}^{-1}\,L_n^{|m|}(\ze)\qq \Psi(-n,1;\ze)=(-1)^n\,n!\,L_n(\ze)\qq n\in{\mb N}\]
give (\ref{eigfcts0}) for the eigenfunctions. $\quad\Box$

Let us point out that the result obtained here for the energies agrees with the result obtained in \cite{behr} for $N=2$. In this reference the authors obtained the quantum energies using for separation variables the cartesian coordinates $(x_1,\,x_2)$. This reflects the superintegrability of this system which allows separation of variables for several different choices of coordinates.

Let us observe that in \cite{behr} the quantization is done in flat space while we have quantized in curved space. Remarkably enough both approaches lead to the same energies while, of course, the eigenfunctions are markedly different. Let us examine the relations between the two approaches.

Starting from formula (\ref{Hspe}) and quantizing in flat space the authors of \cite{behr} obtained
\beq
\frac 12\Big(\wh{P}_1^2+\wh{P}_2^2+\Om^2(x_1^2+x_2^2)
\Big)\Psi(x_1,x_2)=E\,\Psi(x_1,x_2)\qq \Om(E)=\sqrt{\xi-2\rho\,E}\eeq
which is nothing but the sum of two harmonic oscillators. So the energies and eigenfunctions follow easily
\beq
E=(n_1+n_2+1)\Om(E)
\eeq
and solving this relation for $E$ we recover the formula (\ref{spectrum0}) up to the identification $n_1+n_2=n+2|m|$.
The eigenfunctions are expressed in terms of Hermite polynomials which become, using our polar coordinates 
\beq\label{eig2}
{\cal H}_{n_1,n_2}(\ze,\phi)=e^{-\ze/2}\,H_{n_1}(\sqrt{\ze}\,\cos\phi)\,H_{n_2}(\sqrt{\ze}\,\sin\phi). 
\eeq  
The relation between these two bases of the Hilbert space, as shown in Appendix A, is given for $m\geq 0$ by\beq\label{2F1}\barr{l}\dst 
2^{2n+m}\,n!\,\Psi_{n,m}(\ze,\phi)=\sum_{k=0}^n\,i^{k+m} {n \choose k}\,_2F_1\left(\barr{c}-k ,\,-m-n\\n-k+1\earr;-1\right)\,{\cal H}_{k,2n+m-k}(\ze,\phi)\\[4mm]
\dst \hspace{5mm}+\sum_{k=n+1}^{2n+m}\,i^{k+2n+m}{m+n \choose k-n}\,_2F_1\left(\barr{c} k-2n-m ,\,-n\\k-n+1\earr;-1\right)\,{\cal H}_{k,2n+m-k}(\ze,\phi)\earr
\eeq
showing that we have indeed the relation $n_1+n_2=2n+m$.

The relation $\Psi_{n,m}(\ze,\phi)=\Psi^*_{n,|m|}(\ze,\phi)$ gives the corresponding formula for $m<0$.

\section{Action coordinates for $\mathbf{g}_+$}
In proposition (\ref{quinze}) we have seen that in some special cases the geodesics are bounded and closed. This allows us to determine the action coordinates. 

\begin{nth}\label{gplus} For the invariant torus $(H=E,\,P_{\phi}=L>0)$, with $\rho\in\,(0,1)\cup (1,+\nf)$ and $\xi>0$, we have
\beq\label{action2}
I_{\phi}=L\qq I_{\chi}=-L+\sqrt{\xi-2(\rho-1)E}-\sqrt{\xi-2\rho E}\qq E\in\left[E_+,\,\frac{\xi}{2\rho}\right)
\eeq
and the hamiltonian exhibits again degeneracy:
\beq\label{ham2}
H(J)=J\left[\sqrt{\rho(\rho-1)J^2+\xi}-\left(\rho-\frac 12\right)J\right]\qq J\equiv I_{\chi}+I_{\phi}\in\left[L,\sqrt{\frac{\xi}{\rho}}\,\right).
\eeq
\end{nth}

\nin{\bf Proof:} The argument is similar to the one given for the metric $\rg_0$. We have again $I_{\phi}=L$ and it  remains to compute
\[I_{\chi}=\frac 1{2\pi}\oint P_{\chi}\,d\chi=\frac 2{\pi}\int_{\chi_-}^{\chi_+}\ P_{\chi}\,d\chi=\frac 1{\pi}\int_{u_-}^{u_+} \sqrt{\si\,u^2+(2E+L^2)u-L^2}\,\frac{du}{u(1-u)}\]
where $u_{\pm}$, ordered as $u_-<u_+$, are the roots of the polynomial inside the square root.

The first coordinate change
\[\frac 1u=r(1+e\,\cos\tht)\qq r=\frac{E+L^2/2}{L^2} \qq 
e=\sqrt{1-\frac{L^2|\si|}{(E+L^2/2)^2}}<1\]
gives
\[I_{\chi}=\frac{L s e^2}{\pi}\int_0^{\pi}\frac{\sin^2\tht}{(1+e\,\cos\tht)(1+se\,\cos\tht)}\, d\tht \qq\qq s=\frac r{r-1}=\frac{E+L^2/2}{E-L^2/2}>0.\]
Let us notice that 
\[se-1=\frac{\sqrt{\De}-(E-L^2/2)}{E-L^2/2}=\frac{2L^2(\xi-2\rho\,E)}{(E-L^2/2)(\sqrt{\De}+(E-L^2/2)}<0\]
hence both $e$ and $se$ are strictly less than one.

The second coordinate change $t=\tan(\tht/2)$ gives for final result
\[I_{\chi}=\frac{4Le^2 s}{\pi}\int_{-\nf}^{+\nf}\frac{t^2\,dt}{(1+t^2)[1+e+(1-e)t^2][1+se+(1-se)t^2]}\]
which can be computed by the residue theorem and gives (\ref{action2}).
 
Differentiating this relation gives that $D_E\,J>0$ showing that both $J(E)$ and its inverse $E(J)$ are strictly increasing in their respective domains. The computation of $E(J)$ is easily obtained by two successive squarings. $\quad\Box$

Let us determine, for the classical hamiltonian $H_+$ given by (\ref{hamii2}), the discrete spectrum of its quantum extension.

\section{Point spectrum for the hamiltonian on $\mathbf{g}_+$}
Using Carter quantization we have
\beq
\wh{H}_+=-\frac 12\,\Delta+V(\chi) \qq\qq V(\chi)=\frac{\xi\,\sinh^2\chi}{2(1+\rho\,\sinh^2\chi)}\qq \xi>0.
\eeq
An elegant approach was used in \cite{behr} to determine the spectrum of $\wh{H}_0$. As we will see it works also for $\wh{H}_+$.

\subsection{Spectral analysis}
The basic idea is to find coordinates for which the radial Schr\" odinger operator takes the form
\beq
-\frac{d^2}{dQ^2}+V(Q)\eeq
and then use some results given in \cite{gtv}.

The coordinate $Q$, defined as the coordinate conjugate to
\[\Pi=\frac{\cosh(\chi)}{\sqrt{1+\rho\,\sinh^2\chi}}\,P_{\chi},\]
is given by \footnote{It is possible to express $Q$ in terms of elementary functions but this is not useful.}
\beq
Q(\chi)=\int_0^{\chi}\,\frac{\sqrt{1+\rho\,\sinh^2 u}}{\cosh u}\,du\eeq
From which we deduce that the application $\chi\to Q$ is a strictly increasing $C^{\nf}$ diffeomorphism of $(0,+\nf)$ into itself with
\[Q(\chi)=\chi+{\cal O}(\chi^3)\qq\qq Q(+\nf)=+\nf.\]
After the factoring $\Psi(\chi,\phi)=\psi(\chi)\,e^{im\phi}$ the ODE for $\psi(\chi)$ becomes
\beq\label{odeHplus}
-\frac{\cosh^2\chi}{2(1+\rho\,\sinh^2\chi)}\left(\psi''+\frac 1{\tanh\chi}\psi'\right)+V\,\psi=E\,\psi.
\eeq
Since we have for the norm
\[ ||\Psi||^2\propto\int_0^{+\nf}\,\sqrt{1+\rho\,\sinh^2\chi}\tanh\chi\,|\psi(\chi)|^2\,dQ\]
we will define
\beq
\psi(\chi)=(\tanh\chi)^{-1/2}(1+\rho\,\sinh^2\chi)^{-1/4}\,R(\chi)\quad\Rightarrow\quad ||\Psi||^2\propto\int_0^{+\nf}\,|\wti{R}(Q)|^2\,dQ
\eeq
where $\wti{R}=R\circ \chi$.

Transforming the ODE in (\ref{odeHplus}) one obtains
\beq\label{formsym}
\frac 12\left(-\frac{d^2 \wti{R}}{dQ^2}+V_m(Q)\,\wti{R}\right)=E\,\wti{R} \qq V_m=U_m\circ \chi \eeq
with the potential
\beq
U_m(\chi)=\frac{m^2-1/4+1/4\sinh^2\chi}{\tanh^2\chi(1+\rho\,\sinh^2\chi)}+2V(\chi)-\frac{(1-\rho)}{4}\frac{[2+(1-3\rho)\sinh^2\chi-4\rho\,\sinh^4\chi]}{(1+\rho\,\sinh^2\chi)^3}.
\eeq
So we have to consider the formally symmetric operator
\beq
T_m=-\frac{d^2 }{dQ^2}+V_m(Q){\mb I}\qq\qq Q\in\,(0,+\nf)\qq m\in{\mb Z}\eeq
in the Hilbert space $L^2({\mb R}_+)$. Let us prove:

\begin{nth} For all $m\in\,{\mb Z}$ there is a unique  self-adjoint (s.a.) extension of $T_m$ having for spectrum
\beq
\si_{ess}(T_m)=[a,+\nf)\qq\qq\si_{disc}(T_m)\subset [0,a)\eeq
where 
\beq
a=\lim_{Q\to +\nf}\,V_m(Q)=\frac{\wti{\xi}}{2\rho}\qq\qq \wti{\xi}=\xi+\frac 14\qq\qq \xi>0.
\eeq 
\end{nth}

\nin{\bf Proof:} The potential is $C^{\nf}$ on 
${\mb R}_+$ and we have for $Q\to 0+$:
\beq
V_m(Q)=\frac{m^2-1/4}{Q^2}+{\cal O}(1).\eeq
Let us define 
\beq
W_m(Q)=V_m(Q)-\frac{(m^2-1/4)}{Q^2}-a\qq \eeq
which is continuous, bounded and vanishes for $Q\to +\nf$ hence $W_m(Q)\,{\mb I}$ defines a compact operator on $L^2({\mb R}_+)$. We may write
\beq
T_m=t_m+(W_m+a)\,{\mb I}\qq\qq t_m=-\frac{d^2}{dQ^2}+\frac{(m^2-1/4)}{Q^2}
\eeq
where the operator $t_m$ is known as a Calogero hamiltonian which has been thoroughly analyzed in \cite{gtv}[p. 248] where the following results were proved:
\brm
\item The s.a. extension of $t_m$ (hence for $T_m$) is unique for $m\neq 0$. This is not true for $m=0$, since the defect indices are 
$(1,1)$: there is a one parametric $U(1)$ family of self-adjoint extensions. 
\item The essential spectrum (simple and continuous) is:
\beq
\forall m \in{\mb Z}: \qq \si_{ess}(t_m)=[0,+\nf).
\eeq
\erm
Adding a compact operator does not change the essential spectrum so we have
\beq
\si_{ess}(T_m)=\si_{ess}(t_m+a\,{\mb I}+W_m\,{\mb I})=\si_{ess}(t_m+a\,{\mb I})=[a,+\nf).\eeq

For the spectrum positivity some care is needed. For $m\neq 0$ the formal positivity implies the true positivity of its unique s.a. extension. This is no longer true for $m=0$ because we have a one parameter $U(1)$ family of s.a. extensions \cite{gtv}[p. 458] with the following boundary condition at $Q\to 0+$:
\[R_{\la}(Q)=C\Big[\sqrt{Q}\,\ln(k_0 Q)\,\cos\la+\sqrt{Q}\sin\la\Big]+{\cal O}(Q^{3/2}\ln Q)\qq\quad |\la|\leq \frac{\pi}{2}\]
hence for $\chi\to 0+$: 
\[\psi_{\la}(\chi)=C\Big[\ln(k_0 \chi)\,\cos\la+\sin\la\Big]+{\cal O}(\chi\ln\chi).\] 
We will choose the Friedrichs extension (for $|\la|=\pi/2$) with no logarithm and positive spectrum. All the other extensions have a negative mass. 

Hence we will have, for our choice of s.a. extension, that $\si(t_m)\subset {\mb R}^+$  and the positivity of $W_m{\mb I}$ implies $\si(T_m)\subset {\mb R}^+$. So we conclude that $\si_{disc}(T_m)=\si(T_m)\backslash\si_{ess}(T_m)\subset [0,a)$ for all $m\in{\mb Z}$.$\quad \Box$

\vspace{2mm}
\nin{\bf Remark}: the spectral analysis developed here for $H_+$ would be exactly the same as for $H_0$ and refines the results obtained in \cite{behr}. It explains also the apparent degeneracy for $H_0$ of the eigenfunction with  $m=0$: it is related to the non-uniqueness of the s.a. extensions.
   
Let us determine the explicit form of the point spectrum for $\wh{H}_+$.

\subsection{The point spectrum}
\begin{nth} The point spectrum of $\wh{H}_+$ is given by
\beq
E_{n,m}=\wti{J}\Big[\sqrt{\wti{\xi}+\rho(\rho-1)\wti{J}^2}-(\rho-1/2)\wti{J}\Big], \qq \wti{J}=2n+|m|+1, \qq \wti{\xi}=\xi+\frac 14.
\eeq
where $\wti{J}$ is constrained by $\dst\wti{J}<\sqrt{\wti{\xi}/\rho}$, hence there is a finite number of energy levels. The eigenfunctions 
\beq\label{eiggplus}
\Psi_{n,m}(\chi,\phi)=(\tanh\chi)^{|m|}\,(\cosh\chi)^{-1/2-\sqrt{\de}}\,P_n^{(|m|,\sqrt{\de})}(1-2\tanh^2\chi)\,e^{im\phi}\quad \de=\wti{\xi}-2\rho\,E,\eeq
are expressed in terms of the Jacobi polynomials.
\end{nth}

\nin{\bf Proof:} Omitting the intermediate steps already explained when dealing with $\wh{H}_0$ and switching to the variable $u=\tanh^2\chi$, the radial ODE 
\[4u^2(1-u)^2\Psi''+2u(1-u)(2-3u)\Psi'+(\si\,u^2+2E\,u-m^2(1-u))\Psi=0\]
is solved by the change of function
\[\Psi(u)=u^{|m|/2}(1-u)^{1/4+\sqrt{\de}/2}\,R(u)\qq \de=\wti{\xi}-2\rho\,E\qq \wti{\xi}=\xi+\frac 14.\]
The resulting ODE for $R$ is solved by the hypergeometric function
\[ _2F_1(a_-,a_+;|m|+1;u)\qq\qq a_{\pm}=\frac 12(|m|+1+\sqrt{\de}\pm\sqrt{\De})\]
where
\[\De=\wti{\xi}-2(\rho-1)E.\]
The square-summability of the wave function requires now
\[\int_0^1(1-u+\rho\,u)\frac{|R(u)^2|}{(1-u)^{3/2}}\,du<+\nf.\]
For $u\to 0+$ and $m\neq 0$ the second solution of the hypergeometric ODE has for behavior $R(u)\equiv u^{-|m|/2}$ which must be rejected. This is not the case for $m=0$ since then the second linearly independent solution \footnote{The dots just involve an entire function irrelevant for our argument.}
\[_2F_1\left(\barr{c} a_-,\,a_+\\1\earr;u\right)\,\ln u+\ldots\]
exhibits just a harmless logarithmic singularity. However , as explained in section 5.1, we consider the s.a. extension with no logarithm and this function must be rejected.

For $u\to 1-$ the key relation (see \cite{be}[vol. 1, p. 108]) is

\[\barr{l}_2F_1\left(\barr{c} a_- ,\,a_+\\ |m|+1\earr;u\right)=A_2F_1\left(\barr{c} a_- ,\,a_+\\ a_-+a_+-|m|\earr;1-u\right)+\\[6mm]
\hspace{3cm}+B\,(1-u)^{|m|+1-a_--a_+}\,_2F_1\left(\barr{c} |m|+1-a_- ,\,|m|+1-a_+\\ |m|+2-a_--a_+\earr;1-u\right)\earr \]
where
\[A=\frac{\G(|m|+1)\G(|m|+1-a_--a_+)}{\G(|m|+1-a_-)\G(|m|+1-a_+)}\qq\qq B=\frac{\G(|m|+1)\G(a_-+a_+-|m|-1)}{\G(a_-)\G(a_+)}.\]
It shows that the first term is smooth while the second one gives for equivalent
\[R(u)\sim B\,(1-u)^{1/4-\sqrt{\de}/2}\qq\Longrightarrow\qq \frac{|\Psi(u)|^2}{(1-u)^{3/2}}\sim  B^2(1-u)^{-1-\sqrt{\de}}\]
which is never integrable, except if $B=0$. This implies that we must have either $a_+=-n$ for $n\in{\mb N}$, which is excluded since $a_+$ is positive, or $a_-=-n$ which boils down to
\[\wti{J}\equiv 2n+|m|+1=\sqrt{\De}-\sqrt{\de}=\sqrt{\wti{\xi}-2(\rho-1)E}-\sqrt{\wti{\xi}-2\rho E}\qq\quad E\in\left(0,\frac{\wti{\xi}}{2\rho}\right).\]
Since the right hand side is an increasing bijection which maps 
\[E\in\left(0,\frac{\wti{\xi}}{2\rho}\right)\ \to\ \wti{J}\in\left(0,\sqrt{\wti{\xi}/\rho}\right)\qq\Longrightarrow\qq \wti{J}<\sqrt{\wti{\xi}/\rho}\]
giving the required constraint. The inverse function expressing the energy in terms of $\wti{J}$ was already obtained in Proposition \ref{gplus}.

The eigenfunctions obtained can be written
\[ (\tanh\chi)^{|m|}\,(\cosh\chi)^{-1/2-\sqrt{\de}}\,
_2F_1\left(\barr{c} -n ,\,n+|m|+1+\sqrt{\de}\\ |m|+1\earr;\tanh^2\chi\right)\,e^{im\phi}\]
and using the relation with Jacobi polynomials given in \cite{be}[p. 170] we obtain, up to an irrelevant factor, the relation (\ref{eiggplus}). $\quad\Box$

The results obtained here are in perfect agreement with the spectral analysis developed in section (5.1).

\section{Conclusion}
Let us conclude with the following remarks:
\brm
\item We have checked that Koenigs derivation of his SI metrics and the derivation from the framework laid down by Matveev and Shevchishin are in perfect agreement. This last approach leads, in our opinion, to a more elegant  classification involving only three cases: the trigonometric, hyperbolic and affine ones.
\item In the hyperbolic case, as first observed in \cite{kclm}, closed geodesics do appear but only for very special values of the parameters. 
\item The disappointing fact is that all the globally defined systems live on {\em non-compact}  manifolds, namely ${\mb R}^2$ or ${\mb H}^2$. This lack of compact manifolds led Matveev and Shevchishin \cite{ms} to look for generalizations with one linear and two cubic rather than quadratic integrals. As shown in \cite{vds} one obtains cubically SI systems defined on a closed manifold, namely ${\mb S}^2$. In this case a direct analysis \cite{Va3} proves that the metrics are Zoll 
i. e. all the geodesics are closed for all the values taken by the parameters.

A more abstract proof, not relying on the detailed form of the metrics but taking into account the cubic integrals allowed Kiyohara \cite{Ki} to give a different proof of the fact that the metrics must be Zoll.

Another peculiarity of cubically SI models, at variance with Koenigs models, is that no potential is possible  \cite{Ma}.
 
\item Among all of the Koenigs models the one given by equation (\ref{Hspe}) in Subsection 3.1 is somewhat special. Its hamiltonian
\[h=\frac 1{2(1+\rho\,r^2)}\Big(P_1^2+P_2^2+\xi\,r^2\Big)\qq\qq r^2=x_1^2+x_2^2\]
was generalized quite recently by Ra$\tilde{\rm n}$ada \cite{Ra} to  
\[\wti{H}_a(\ka=-\rho,\alf^2=\xi)=h+\frac 1{(1+\rho\,r^2)}\left(\frac{k_1}{x_1^2}+\frac{k_2}{x_2^2}\right),\]
still quadratically SI but not globally defined since the new potential is singular at the origin. 
\item As shown in \cite{Va1}, the same hamiltonian with a different potential:
\beq\left\{\barr{l}\dst 
H=h+\frac{(-2\,\rho\,l\,x_1+m)}{2(1+\rho\,r^2)} \\[5mm]
Q=2H\,L_3+l\,P_2,\earr\right.\eeq
gives a cubically {\em integrable} system.
\item Changing again the potential, as shown in \cite{Va2}, we have  
\beq\left\{\barr{l}\dst 
H=h+\frac{(-\rho\,k(x_1^2+x_2^2)-2\,\rho\,l\,x_1+m)}{2(1+\rho\,r^2)}\\[5mm]  
Q=2H\,L_3^2+k\,L_3^2+2l\,P_2\,L_3+l^2\,x_2^2.\earr\right.\eeq
which is a quartically {\em integrable} system. Quite unexpectedly the {\em same} metric, globally defined on $\,M\cong{\mb R}^2$, when subjected to a change of its potential, may lead either to SI or to integrable systems with integrals of various degrees 
in the momenta. Is this phenomenon commonplace or exceptional?
\erm

\vspace{5mm}
\nin {\bf Acknowledgements:} We would like to thank Philippe Briet for his kind help with the spectral analysis of Section 5.1. 

\renewcommand\thesection{Appendix \Alph{section}}
\numberwithin{equation}{section}

\appendix

\section{Relation between two bases}
The two bases $\Psi_{n,m}$ and ${\cal H}_{n_1,n_2}$ are defined in relations (\ref{eigfcts0}) and (\ref{eig2}). Since they are orthogonal we must have the expansion
\beq
\Psi_{n,m}(\ze,\phi)=\sum_{n_1,n_2\geq 0}\,c^{n_1,n_2}_{n,m}\ {\cal H}_{n_1,n_2}(\ze,\phi),
\eeq
where the coefficients, using an orthogonality relation, are given by
\beq
2^{n_1+n_2}\,c^{n_1,n_2}_{n,m}=\int_0^{+\nf}\,\int_0^{2\pi}\,\frac{{\cal H}_{n_1,n_2}(\ze,\phi)}{n_1!\,n_2!}\,\Psi_{n,m}(\ze,\phi)\,d\ze\,d\phi.
\eeq
Using the generating function of the Hermite polynomials
\beq
\sum_{n\geq 0}\frac{\la^n}{n!}\,H_n(x)=e^{-\la^2+2\la\,x}
\eeq
we will compute
\beq\label{sumH}
S\equiv\sum_{n_1,n_2\geq 0}\la^{n_1}\,\mu^{n_2}\,2^{n_1+n_2}\,c^{n_1,n_2}_{n,m}\eeq
given by
\beq
S=e^{-\la^2-\mu^2}\int_0^{+\nf}\,e^{-\ze}\ze^{|m|/2}L_n^{|m|}(\ze)\int_0^{2\pi}\,e^{im\phi}\,e^{2\la\sqrt{\ze}\cos\phi+2\mu\sqrt{\ze}\sin\phi}\,\frac{d\phi}{2\pi}\,d\ze.
\eeq
The $\phi$ integral, setting $z=e^{i\phi}$, becomes
\beq
\frac 1{2\pi i}\oint\frac{dz}{z}\,z^m\,e^{(\la-i\mu)\sqrt{\ze}\,z+(\la+i\mu)\sqrt{\ze}/z}
\eeq
where the contour is the circle of radius one. The residue theorem gives, for $m\geq 0$:
\beq
S=\sum_{k\geq 0}\frac{(\la-i\mu)^k}{k!}\frac{(\la+i\mu)^{k+m}}{(k+m)!}e^{-\la^2-\mu^2}\,\int_0^{+\nf}\,e^{-\ze}\ze^{k+m}L_n^{m}(\ze)\,d\ze.
\eeq
This integral is computed using the Rodrigues formula for Laguerre polynomials and one obtains
\beq
\int_0^{+\nf}\,e^{-\ze}\ze^{k+m}L_n^{m}(\ze)\,d\ze=\left\{\barr{ll} 0 & \qq k\leq n-1\\[4mm]\dst 
\frac{k!}{(k-n)!}\,\frac{(m+k)!}{n!} & \qq k\geq n\earr\right.\eeq
and the remaining sum does factorize to
\beq
S=\frac{(\la-i\mu)^n}{n!}\,(\la+i\mu)^{n+m}.\eeq
Its value for $m<0$ is merely obtained by complex  conjugation.

We need to expand this function in powers of $\la$ and $\mu$. The binomial theorem gives
\beq
n!\,S=\sum_{k=0}^{m+n}\sum_{l=0}^n\,i^{m-k+l}{n \choose l}{m+n \choose k}\la^{k+l}\,\mu^{2n+m-(k+l)}
\eeq
and the change of summation index $l=\nu-k$, followed by an interchange of the summations, allows to write $S=S_1+S_2$ with
\beq\barr{ll}\dst
n!\,S_1=\sum_{\nu=0}^n i^{\nu+m}\la^{\nu}\,\mu^{2n+m-\nu}\sum_{k=0}^{\nu}(-1)^k{m+n \choose k}{n \choose \nu-k},\\[5mm]\dst
n!\ S_2=\sum_{\nu=n+1}^{2n+m} i^{\nu+m}\la^{\nu}\,\mu^{2n+m-\nu}\sum_{k=0}^n(-1)^{n-k}{m+n \choose \nu-n+k}{n \choose n-k}.\earr\eeq
It is convenient to use Pochammer symbols defined by
\[(a)_0=1 \qq\qq (a)_n=a(a+1)\cdots (a+n-1)\quad n\geq 1\] 
and the identities  
\beq
(-n)_k=(-1)^k\frac{n!}{(n-k)!}\quad k\leq n\qq\qq (n)_{k+l}=(n)_k\,(n+k)_l
\eeq
to get
\beq
\sum_{k=0}^{\nu}\,(-1)^k{m+n \choose k}{n \choose \nu-k}={n \choose \nu}\sum_{k=0}^{\nu}(-1)^k\frac{(-m-n)_k\,(-\nu)_k}{k!\,(n-\nu+1)_k}.\eeq
This last sum, expressed with Gauss hypergeometric function \cite{be}[vol. 1, p. 56], gives eventually 
\beq 
n!\,S_1=\sum_{\nu=0}^n i^{\nu+m}\,{n \choose \nu}\,  _2F_1\left(\barr{c} -\nu ,-m-n\\n-\nu+1\earr;-1\right)
\la^{\nu}\,\mu^{2n+m-\nu}.
\eeq
The computation of $S_2$ is similar. The relation
\beq
(-1)^{n-k} {m+n \choose \nu-n+k}=(-1)^n {m+n \choose \nu-n}\frac{(\nu-2n-m)_k}{(\nu-n+1)_k}\eeq
gives
\beq
\sum_{k=0}^n\,(-1)^{n-k}{m+n \choose \nu-n+k}{n \choose n-l}=(-1)^n {m+n \choose \nu-n}\,_2F_1\left(\barr{c} \nu-2n-m ,-n\\ \nu-n+1\earr;-1\right)\eeq
from which we conclude to
\beq
n!\,S_2=\sum_{\nu=n+1}^{2n+m}i^{\nu+2n+m}{m+n \choose \nu-n}\,_2F_1\left(\barr{c} \nu-2n-m ,-n\\ \nu-n+1\earr;-1\right)\,\la^{\nu}\,\mu^{2n+m-\nu}.
\eeq
Having computed $S=S_1+S_2$ and comparing the powers of $\la$ and $\mu$  with (\ref{sumH}) ends up the proof of (\ref{2F1}). $\quad\Box$


\end{document}